\documentclass[10pt, sigconf,dvipsnames]{acmart}

\acmConference{}{}{}
\acmBooktitle{}
\acmYear{}
\acmISBN{}
\acmDOI{}

\renewcommand\footnotetextcopyrightpermission[1]{}

\AtBeginDocument{%
  
}

\usepackage{geometry}

\usepackage{graphicx}
\usepackage{subcaption}

\usepackage{amsmath,mathtools}
\usepackage{bbding}
\usepackage{wasysym}
\usepackage{float}
\usepackage{adjustbox}

\usepackage[ruled,lined,linesnumbered]{algorithm2e}
\SetAlgoCaptionLayout{centerline}

\usepackage{enumitem}
\usepackage{booktabs}

\usepackage{gensymb}


\begin{document}

\title{LEO Topology Design Under Real-World Deployment Constraints}

\author{Muaz Ali}
\affiliation{\institution{University of Arizona} \country{Tucson, AZ, USA}}

\author{Beichuan Zhang}
\affiliation{\institution{University of Arizona} \country{Tucson, AZ, USA}}

\begin{abstract}

The performance of large-scale Low-Earth-Orbit (LEO) networks, which consist of thousands of satellites interconnected by optical links, is dependent on its network topology. Existing topology designs often assume idealized conditions and do not account for real-world deployment dynamics, such as partial constellation deployment, daily node turnovers, and varying link availability, making them inapplicable to real LEO networks. In this paper, we develop two topology design methods that explicitly operate under real-world deployment constraints: the Long--Short Links (LSL) method, which systematically combines long-distance shortcut links with short-distance local links, and the Simulated Annealing (SA) method, which constructs topologies via stochastic optimization. Evaluated under both full deployment and partial deployment scenarios using 3-months of Starlink data, our methods achieve up to 45\% lower average end-to-end delay, 65\% fewer hops, and up to $2.3\times$ higher network capacity compared to \texttt{+Grid}. Both methods are designed to handle daily node turnovers by incrementally updating the topology, maintaining good network performance while avoiding costly full reconstruction of the topology.

\end{abstract}

 \maketitle

\section{Introduction}
\label{sec:Introduction}

Operating at altitudes below 2,000 km, Low-Earth-orbit (LEO) constellations are rapidly expanding, with operators like Starlink having already deployed thousands of satellites to provide global broadband access, and many other companies and countries are planning or deploying their own constellations~\cite{starlink_availability_map,saxena2025starlink8million,deployment_plan_200k}. 
Modern satellites are equipped with laser terminals that enable inter-satellite links (ISLs) to interconnect the satellites forming an in-space network~\cite{starlink_technology,tesat_scot80,starlink_report}. As the LEO constellations grow, designing efficient network topologies becomes increasingly critical to ensure low latency and high capacity for users worldwide.


The most common topology is \texttt{+Grid}, in which each satellite connects to four nearest neighbor satellites: two in the same orbital plane (ahead and behind), and two in adjacent orbital planes (left and right)~\cite{delyHandley,IPCGrid}. \texttt{+Grid} is attractive due to its simplicity and stability. However, because it uses only short links, long-distance routes require many hops, leading to higher end-to-end delay and lower total network capacity. 

Given the wireless nature of the ISLs, a satellite can potentially connect to any satellite within range and line-of-sight, which opens up a much larger design space. In particular, longer links can act as shortcuts that reduce hop count and end-to-end delay. Motivated by this insight, recent work has proposed alternative topology designs, including structured inter-plane patterns (motif-style topologies)~\cite{motif} and optimization-based approaches that explicitly target delay and capacity~\cite{ron2025timedependent,mcts}.
These approaches all improve hop counts and delay compared to \texttt{+Grid}.


\begin{figure}[ht]
  \centering
  \captionsetup[subfigure]{subrefformat=parens, font=scriptsize}

  \begin{subfigure}{0.48\linewidth}
    \centering
    \includegraphics[width=\linewidth]{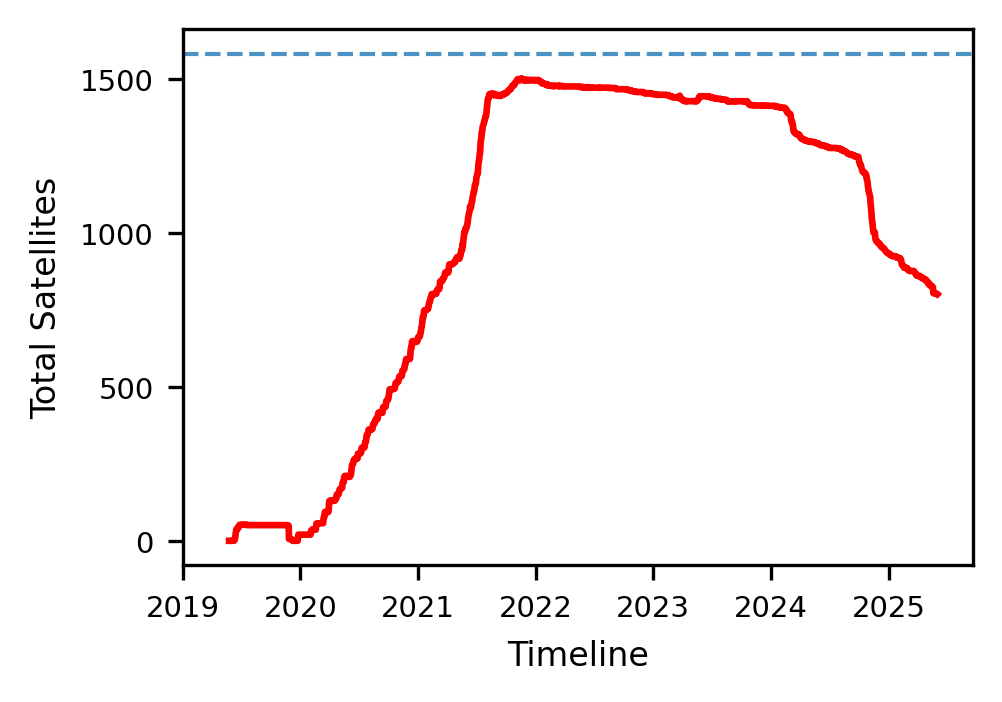}
    \caption{Operational Satellite Count}
    \label{fig:sats_over_time}
  \end{subfigure}\hfill
  \begin{subfigure}{0.49\linewidth}
    \centering
    \includegraphics[width=\linewidth]{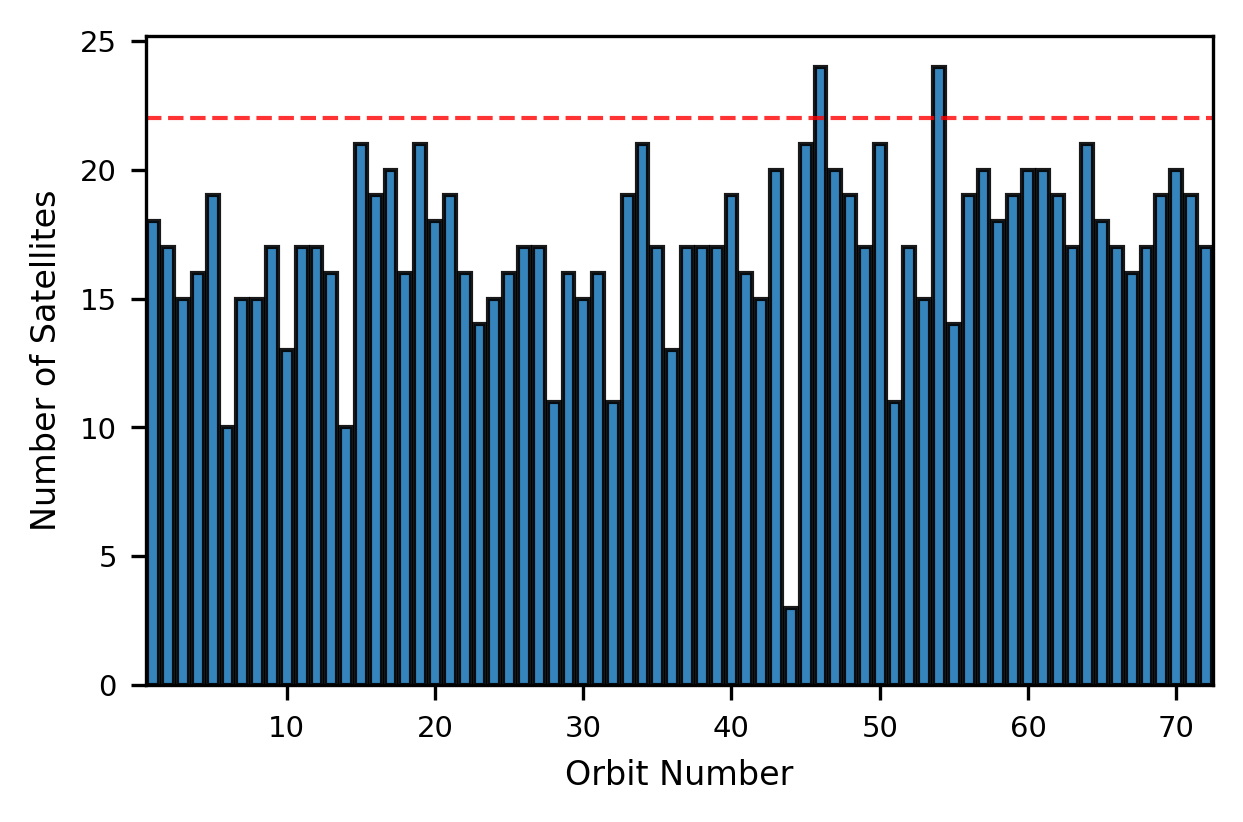}
    \caption{Satellite Across Orbits, 2024-10-01.}
    \label{fig:raan_group_population}
  \end{subfigure}
\caption{Deployment of Starlink Shell-1: partial, uneven, and dynamic. The full deployment would be $72 \times 22 = 1,584$ satellites (dashed line).}
\label{fig:real_asymmetry}
\end{figure}

However, prior work assumes an idealized constellation and does not account for issues arising from real-world deployment. For example, some designs assume full deployment and rely on the symmetry of the constellation in order to apply a repeated pattern across the entire network~\cite{motif}, while others optimize for a given snapshot of the constellation and do not consider the continuous changes of the constellation over days and weeks~\cite{ron2025timedependent,mcts}. These assumptions are common in the literature, but they do not hold in practice. Figure~\ref{fig:real_asymmetry} shows the entire deployment history of Starlink Shell-1. Even though the target is to have 72 orbital planes with 22 satellites each, full deployment has never been achieved since the initial launch in 2019. The number of operational satellites changes continuously as new satellites are launched and existing satellites are moved or de-orbited (Figure~\ref{fig:sats_over_time}). If we zoom in on a particular day, the number of satellites per orbit is also uneven, creating persistent asymmetry (Figure~\ref{fig:raan_group_population}). Other shells of Starlink constellation also exhibit similar behaviors of partial, uneven, and dynamic deployment. It is possible that LEO constellations will eventually stabilize, reaching full deployment and symmetry, but that process will likely take years, and as Figure~\ref{fig:real_asymmetry} shows, there is no sign of such stabilization yet.  As a result, topologies that are not designed with these issues in mind will not be applicable to real-world LEO constellations.


A well-designed topology should have a balanced mix of short and long links to achieve low delay and low hop count, and it must also be robust to the realities of real-world deployment. First, it must handle changing node availability and not assume full deployment or symmetry. Second, it must account for link availability, which may change continuously as satellites move and their relative geometry changes accordingly. Third, it must support incremental updates over time to minimize topology changes while maintaining good performance.


We develop two efficient topology design methods that operate directly on real deployed constellations. The first method, \emph{Long--Short Links (LSL)}, systematically combines short local links with long-distance shortcuts to reduce both delay and hop count. The second method, \emph{Simulated Annealing (SA)}, uses stochastic optimization to search the design space and construct high-performing topologies. It is tunable and can generate a family of designs that trade off delay and hop count. To handle time-varying link availability, both methods only use \emph{stable links}, defined as links that remain available throughout the entire orbital period. To address varying node availability, we introduce an algorithm that incrementally updates the topology as nodes leave or join. This algorithm offers two key benefits: (i) it avoids the costly full reconstruction of the topology, and (ii) it reduces link churn that would otherwise result from building a fresh topology. Together, these components enable effective topology designs under real-world deployment constraints.

We evaluate the new designs using both real deployment data from Starlink Shell-1 (October--December 2024) and synthetic constellations based on Starlink and Amazon Kuiper. On the synthetic, fully deployed Starlink Shell-1, LSL achieves the lowest delay, while SA achieves the lowest hop count, reducing hops by 64\% relative to +Grid and by 60\% relative to Motif.  Under shortest-path routing with max--min fair allocation, LSL delivers about $2\times$ higher aggregate throughput than \texttt{+Grid} and Motif. On real deployed Starlink Shell-1, our methods reduce end-to-end delay by up to 45\%, hop count by up to 65\%, improve throughput by up to $2.3\times$, and maintain low link breakage rates (1.0--1.3\%). Overall, these results demonstrate strong performance under realistic, dynamically evolving LEO deployments.

The rest of the paper is organized as follows. Section~\S\ref{sec:background} reviews background and key concepts. Section~\S\ref{sec:problems_topology_design} discusses the rationale behind LEO topology design. Section~\S\ref{sec:sys_design} presents the two approaches, LSL and SA. Section~\S\ref{sec:evaluation} evaluates the proposed approaches. Section~\S\ref{sec:discussion} discusses implications and limitations, Section~\S\ref{sec:related_works} surveys related work, Section~\S\ref{sec:ethics} reviews ethical considerations, and Section~\S\ref{sec:conclusion} concludes.

\section{Background on LEO Networks}
\label{sec:background}



A satellite moves along an \textit{orbit} around the Earth. For typical LEO satellites, the orbit is nearly circular, and is mainly characterized by its altitude (distance above the Earth's surface), inclination (angle between the equatorial plane), and the Right Ascension of the Ascending Node (RAAN), which can be thought of as the longitude of the point where the satellite crosses the equatorial plane from south to north. Orbits with the same altitude and inclination but different RAAN values are grouped into a \textit{shell}. 
For example, the most studied shell is Starlink Shell-1, which is provisioned for 72 orbits at 550km altitude and 53\degree\ inclination, and 22 satellites per orbit. A LEO constellation typically consists of multiple shells for global coverage and capacity.

An inter-satellite link (ISL) is an optical point-to-point link that connects a pair of satellites and enables direct data transfer between them. 
ISLs are subject to some physical limitations. It is only possible when two satellites are in line-of-sight, and its performance will greatly degrade if the link goes through the atmosphere, which is why ISLs should be at least 80km above the Earth's surface. Within its designed range, an ISL can maintain its peak bandwidth regardless of the inter-satellite distance, but beyond the designed range, the link bandwidth will drop as the distance increases, until it eventually becomes infeasible to maintain the link. It also takes time for two moving satellites to acquire and establish an ISL, which can range from seconds to minutes depending on the hardware and the relative motion of the satellites. 

Despite these limitations, recent LEO constellations demonstrate that high-capacity ISLs are feasible over long distances. Starlink reports optical ISLs capable of sustaining data rates between 100 and 200~Gbps at separations of up to 5{,}400~km, with link acquisition times on the order of seconds and some links remaining stable for weeks~\cite{starlink_report,starlink_technology}. Similarly, Tesat, an ISL terminal provider, reports operational links at 10~Gb/s over distances of up to 8{,}000~km, with designs that are scalable to 100~Gb/s~\cite{tesat_scot80}. 
In this work, we abstract away the physical-layer details by assuming a constant per-link data rate for any feasible ISL whose length does not exceed a fixed maximum distance, for example approximately 8{,}000~km, consistent with current and near-term ISL capabilities~\cite{tesat_scot80}.



Each satellite is equipped with a fixed number of ISL terminals, which limits the number of simultaneous ISLs it can maintain. While early research assumes five ISLs per satellite~\cite{delyHandley}, recent work are all based on the assumption of four ISLs per satellite~\cite{motif,ron2025timedependent,mcts}. Starlink’s website states each satellite can support 3 ISLs~\cite{starlink_technology}. We evaluate our work under 4-ISL constraints in order to compare with prior work, but we also evaluate under 3-ISL constraints for more realistic scenarios.

\begin{figure}[h ]
  \centering
  \captionsetup[subfigure]{font=scriptsize,justification=raggedright,singlelinecheck=false}

  \makebox[\linewidth][c]{%
    \begin{subfigure}{0.30\linewidth}
      \centering
      \includegraphics[width=\linewidth]{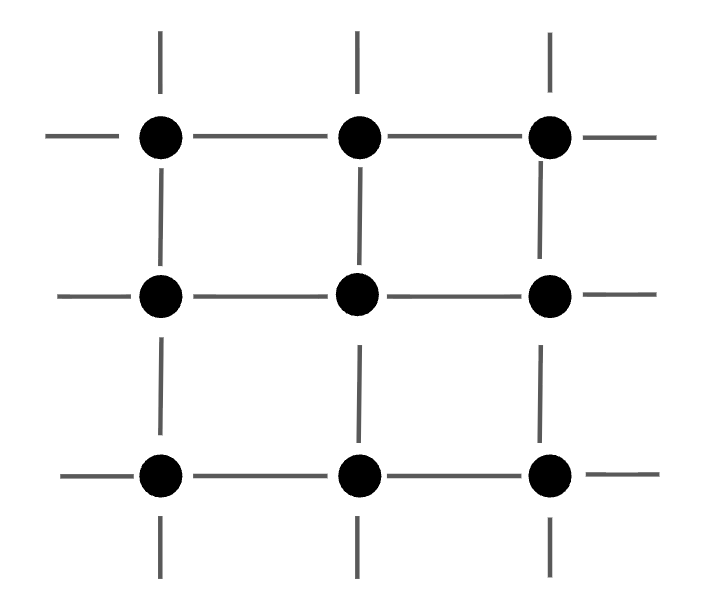}
      \caption{4-ISL grid (+Grid)}
      \label{fig:grid_topology}
    \end{subfigure}
    \hspace{0.06\linewidth}
    \begin{subfigure}{0.30\linewidth}
      \centering
      \includegraphics[width=\linewidth]{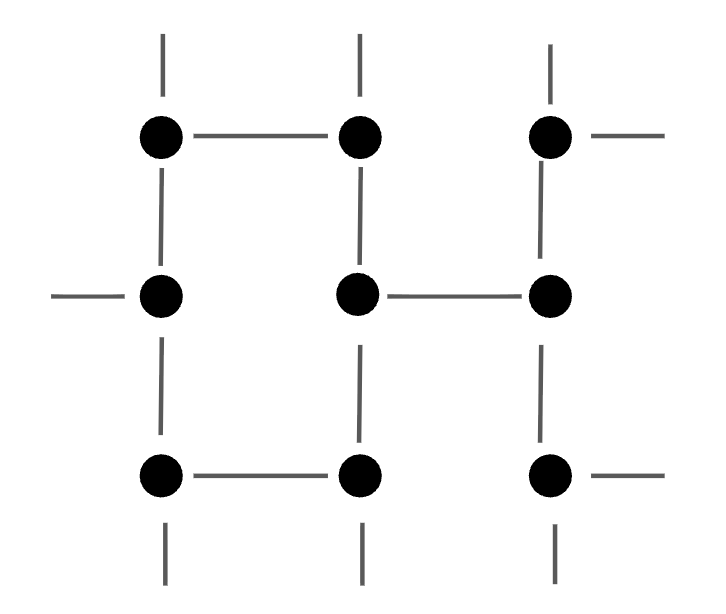}
      \caption{3-ISL Grid}
      \label{fig:3isl_grid_topology}
    \end{subfigure}%
  }

  \caption{Grid Topologies with Different ISL limits}
  \label{fig:grid_and_3isl_grid}
\end{figure}

When each satellite has 4 ISLs per satellite, the most widely used topology for LEO network is the \emph{+Grid} topology~\cite{delyHandley}. It builds a regular mesh by connecting each satellite to:
(i) its two immediate \emph{intra-orbit} neighbors (forward/backward along the same orbital), and
(ii) its nearest \emph{inter-orbit} neighbors in the two adjacent orbits, as illustrated in Figure~\ref{fig:grid_topology}.

When each satellite has only 3 ISLs, the grid can be built by keeping the same intra-orbit links and removing one of the inter-orbit links. 
Instead of having each satellite connect to \emph{both} adjacent orbits, 3-ISL Grid connects inter-orbit links \emph{alternatively}, as illustrated in Figure~\ref{fig:3isl_grid_topology}. 




\section{The Problem of LEO Topology Design}
\label{sec:problems_topology_design}



\begin{figure}[ht]
  \centering
  \includegraphics[width=0.5\linewidth]{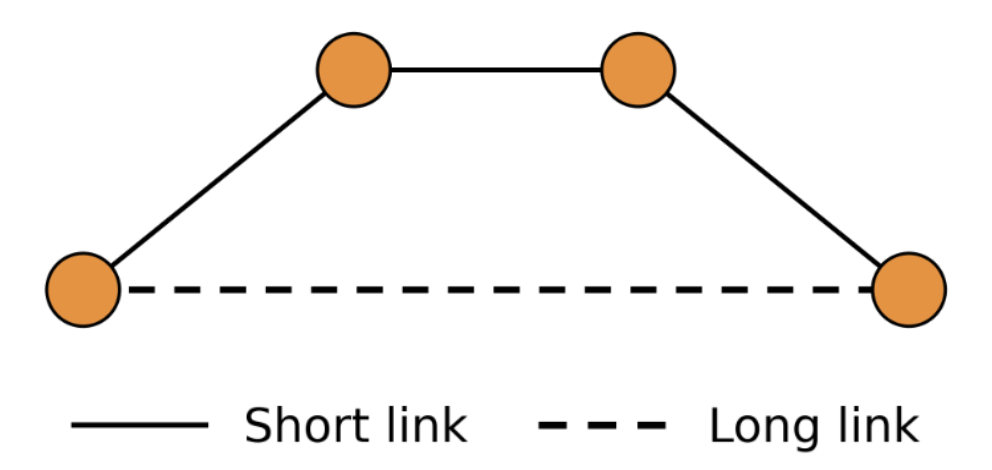}
  \caption{Illustration of short links and a long link. Short links offer local connectivity, while the long link acts as a shortcut.}
  \label{fig:short-long-links}
\end{figure}

A natural starting point for LEO networks is the +Grid topology: it is simple to construct and tends to remain feasible as satellites move. Although ISL distances vary over time, the local links used by grid topologies generally stay within range. For example, in Starlink Shell-1 under full and even deployment, the maximum intra-orbital distance is about \(1{,}950\)~km and the maximum inter-orbital distance is about \(600\)~km, both well within the 8{,}000~km ISL range limit~\cite{tesat_scot80}. The downside, however, is that the exclusive use of short links leads to high hop counts. For example, in the perfect Starlink Shell-1, the worst hop count is 47. When traffic needs to traverse many hops to reach its destination, it takes up more interfaces along the path, which lowers the aggregate network capacity. It also increases end-to-end delay because multi-hop routes may detour around the Earth instead of taking more direct paths (Figure~\ref{fig:short-long-links}).

Since ISLs are wireless, a satellite may connect to any other satellite within its optical range. This opens up the design space to a wide variety of topologies that can exploit links of different lengths. Long links across multiple orbits can reduce hop count and delay, and free up interfaces in the middle. But relying primarily on long links can also perform poorly by weakening local connectivity and forcing indirect paths to nearby destinations. Thus, good network performance requires a \emph{balanced mix} of short links for local structure and long links as shortcuts. 

LEO network topology design can be formulated as a multi-objective optimization problem that jointly minimizes delay, hop count, and link breakage subject to deployment and degree constraints. 
To remain relevant to real-world operations, it is critical to understand the major constraints stemmed from real-world deployments.





\textbf{Node Availability:}
Most prior work assumes fully and evenly deployed constellations as they were provisioned. Real constellations do not look like this. 
It takes years or even a decade to fully deploy a shell, and the active set of satellites is in constant flux as new satellites are launched and existing satellites de-orbit. It is also not uncommon for operators to change their plans in the middle of deployment. As shown in Figure~\ref{fig:real_asymmetry}, the Starlink Shell-1 was initially deployed towards the provisioned 1584 satellites, but the plan changed around the beginning of 2022, and since then the number of satellites in the shell has been declining. In fact, Starlink recently announced plans to lower roughly 4{,}400 satellites from about 550\,km to 480\,km over the course of 2026 to reduce collision risk~\cite{lowering_orbit}.
Therefore a practical topology design must expect partial and uneven deployments, and remain effective as the constellation evolves.


\textbf{Daily Turnover:}
The set of active satellites changes almost every day. Besides new launches and de-orbits, satellites may also be moved out of or into the shell, and may encounter failures that cause them to go offline. 
Depending on the topology design methods, some may have to rerun the optimization every day and end up with a very different topology from the previous day, which can cause network instability and operational complexity. A practical topology design should therefore have an incremental update procedure that accounts for daily node turnover and minimizes unnecessary link changes. 

\textbf{Topological Asymmetry:}
In a fully deployed shell, symmetry makes one snapshot broadly representative of the day. Consider a multi-hop satellite path connecting Los Angeles and London. Even the path length changes as the satellites move, after a few minutes when the next set of satellites moves into position, the path will look similar again due to topological symmetry of the constellation. In contrast, under partial deployment, the path structure can change substantially as satellites move. The same two locations may have a short path at one time and a much longer path hours or days later. A practical topology design must therefore look beyond a single snapshot and evaluate topologies over long time horizons to ensure that they remain effective as the constellation evolves.
For this reason, we decide to evaluate topologies over \emph{all source--destination satellite pairs} rather than optimizing for any particular ground-to-ground traffic matrix.

\begin{figure}[ht]
  \centering
  \begin{subfigure}{0.49\linewidth}
    \centering
    \includegraphics[width=\linewidth]{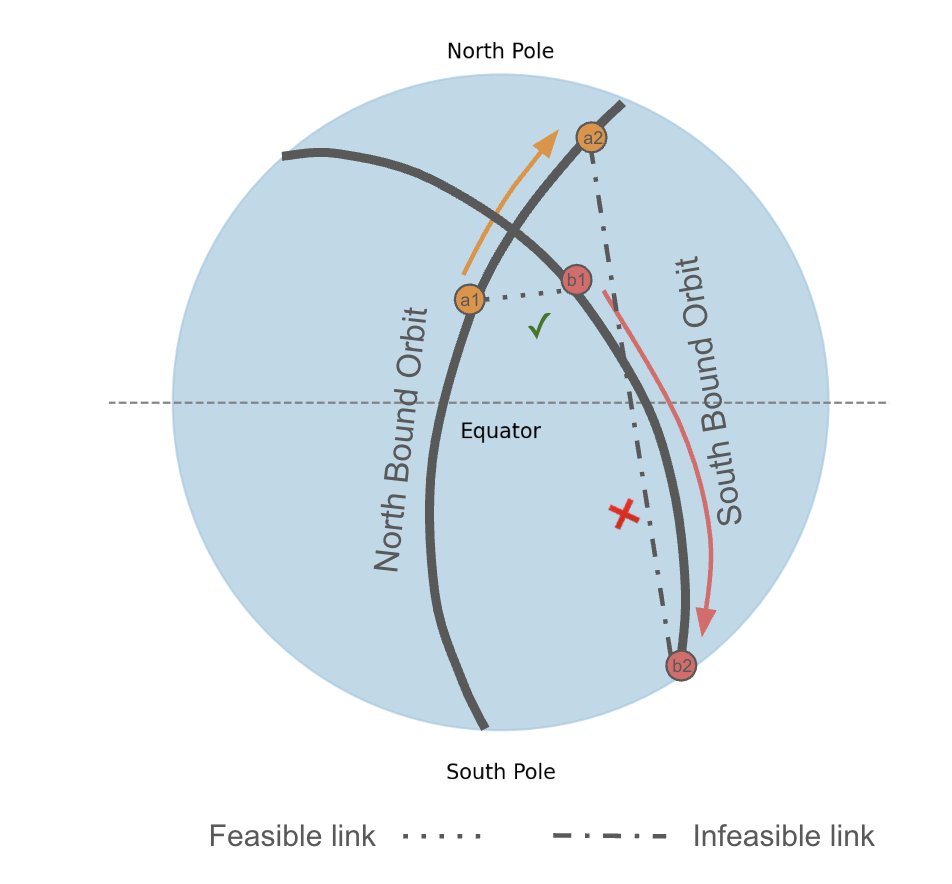}
    \caption{\textbf{Transient feasibility.} Opposite-direction pairs (northbound vs.\ southbound) have feasible link at close approach, then lose line-of-sight as separation increases.}
    \label{fig:link_feasibility_transient}
  \end{subfigure}
  \hfill
  \begin{subfigure}{0.49\linewidth}
    \centering
    \includegraphics[width=\linewidth]{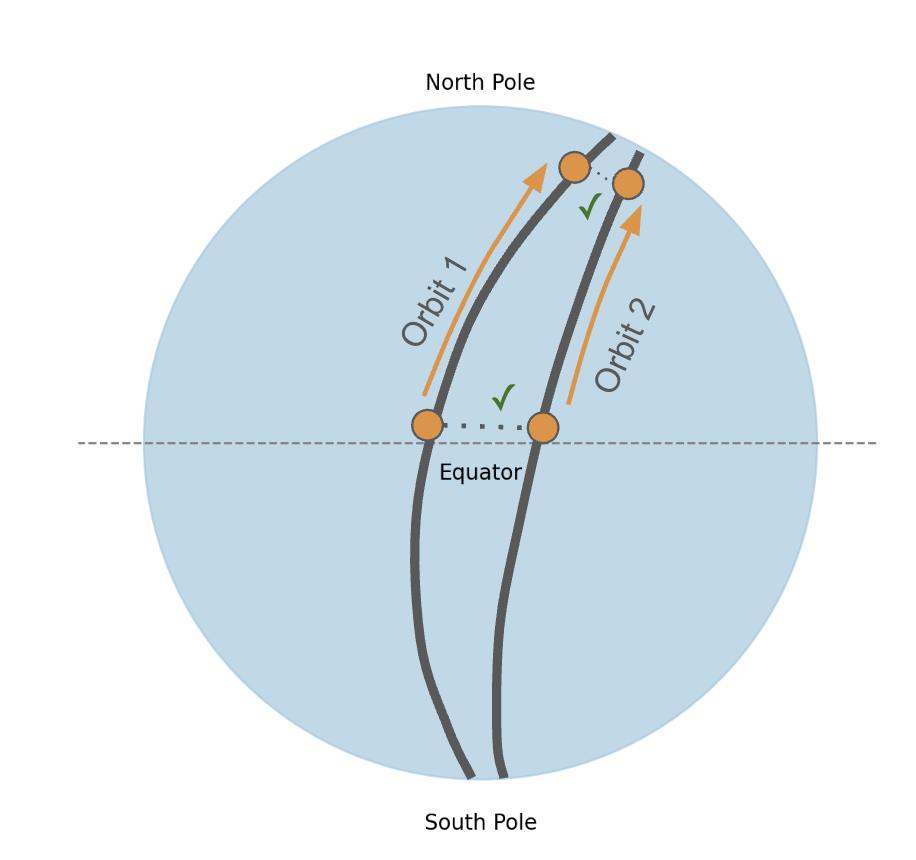}
    \caption{\textbf{Stable feasibility.} Satellites on nearby orbits are farthest near the equator and closest near the poles. The link is stable because \emph{maximum} separation stays within range.}
    \label{fig:link_feasibility_stable}
  \end{subfigure}
\caption{Link feasibility and stability in inclined shells. Opposite-direction pairs are feasible but short-lived and unstable (a), while nearby orbit pairs remain stable because their maximum separation stays within range limits (b).}
\label{fig:link_feasibility}
\end{figure}

\textbf{Link Availability:}
An inter-satellite link (ISL) is \emph{feasible} only when the two satellites are within the maximum optical range and maintain line-of-sight (LoS) above the atmospheric layer~\cite{chaudhry2021laserintersatellitelinksstarlink}, i.e., the LoS path remains at least 80~km above Earth's surface. 
Because inter-satellite separation changes over the orbital cycle, a link that is feasible at one point may violate the range or LoS constraint later. This becomes a problem when we want to incorporate long links into the topology. 
Figure~\ref{fig:link_feasibility_transient} illustrates a rather extreme case. A satellite that is moving northbound and another satellite that is moving southbound can be very close to each other to establish a link at close approach, but as they move apart, the link quickly becomes out-of-range.  




In another example illustrated in Figure~\ref{fig:link_feasibility_stable}, two satellites in nearby but not adjacent orbits move in similar directions. Their distance varies over the orbital cycle: they are farthest apart near the equator and closest near the poles. If their maximum separation remains within the range/LoS limits, we can call this link \emph{stable} because it is feasible at all times and straightforward to maintain. 


Any LEO topology design must be explicit about the types of links it uses. Some prior work considers all links in a snapshot as equally usable, without distinguishing between stable and transiently feasible links, which can lead to topologies that last only for a short time. 
The ideal topology design should be able to exploit links of different levels of availability. In this paper, however, we take a conservative approach of using only stable links, such that we can minimize link churn and avoid making any assumptions about the hardware's ability to acquire and maintain transient links. We leave the design of topologies that also exploit transiently feasible links as future work.

\section{System Design}
\label{sec:sys_design}

Our system consists of three stages: (1) Given a set of satellite nodes, we construct a \emph{set of stable links} that remain feasible through a full orbital cycle. This step explicitly addresses the challenge of link availability. (2) We generate a topology from this stable-link set using one of the two methods: \emph{Long--Short Links (LSL)} or \emph{simulated annealing (SA)}. These methods are designed to produce topologies that achieve low delay and low hop count while respecting degree constraints; they do not assume full deployment or any topological symmetry. (3) We apply an \emph{incremental update} procedure that preserves as much of the previous day's topology as possible while adding or removing links necessary to account for daily node turnover.


\subsection{Stable Links}

\textbf{Definition:}
We define \emph{stable links} as ISLs that remain feasible over a full orbital cycle. Feasibility requires (i) \emph{line-of-sight} (LoS) above an atmospheric layer of height $a$ (e.g., $a=80$\,km) and (ii) within a hardware range limit $D_{\max}$. Because satellites in the same shell have the same altitude, the LoS constraint reduces to a distance bound between two satellites~\cite{chaudhry2021laserintersatellitelinksstarlink}. Let $h$ be the orbital altitude and $r$ be the Earth's radius, the maximum separation between two satellites that satisfy requirement (i) is
\begin{equation}
D_{\mathrm{LoS}} \;=\; 2\sqrt{(r+h)^2 - (r+a)^2}.
\label{eq:dlos}
\end{equation}

Let $d_{x,y}(t)$ be the instantaneous Euclidean distance between satellites $x$ and $y$ at time $t$ over one orbital period $T$, link $(x,y)$ is stable if its \emph{worst-case} separation satisfies both requirements:
\begin{equation}
\max_{t \in [0,T]} d_{x,y}(t) \;\le\; D_{\mathrm{stab}},
\qquad
D_{\mathrm{stab}} \;=\; \min\!\bigl(D_{\max},\, D_{\mathrm{LoS}}\bigr).
\label{eq:stable_def}
\end{equation}


\textbf{Orbital Geometry:}
To find out the largest distance between any two satellites as they move, we need some basic orbital geometry. 

Satellites are usually described by orbital parameters, not Euclidean coordinates. Within a shell, each orbit is identified by its RAAN, denoted by $\Omega$, which can be thought of as the longitude of the orbit.
RAAN ranges from $0^\circ$ to $360^\circ$ around the Earth in the East-West direction. Within an orbit, each satellite can be identified by its \textit{anomaly}, denoted by $u$, which ranges from $0^\circ$ to $360^\circ$ around the Earth in the South-North direction. Anomaly represents the satellite's angular position along its orbit, measured from a common reference point. The pair $(\Omega, u)$ therefore identifies a satellite's position within the shell. 

The angular offset between two satellites can be expressed as $(\Delta\Omega, \Delta u)$, where $\Delta\Omega$ is the difference in RAAN and $\Delta u$ is the difference in anomaly. Larger $\Delta\Omega$ means the two satellites are in more distant orbital planes, while larger $\Delta u$ means they are more out of phase within their respective orbits.
Since two satellites in the same shell move at roughly the same speed, even through their Euclidean distance will change as they move, their angular offset, $(\Delta\Omega, \Delta u)$, remains constant. 

\textbf{Compute Stable-Link Set:}
For a given pair of satellites, we can easily get their angular offset from their orbital parameters. Having the angular offset $(\Delta\Omega, \Delta u)$, we want to know if the link's worst-case distance is within the stable-link threshold $D_{\mathrm{stab}}$.



First, we discretize the entire orbital space into a grid, with one degree resolution in both RAAN and anomaly. Second, we fix satellite A at one point and enumerate satellite B at each of the $360 \times 360$ points and compute the Euclidean distance between A and B. Third, we move satellite A along its orbit by one degree, and repeat the same enumeration and distance computation for satellite B. This process is repeated until satellite A completes a full orbit (360 degrees).  

\begin{figure}[h]
    \centering
    \includegraphics[width=0.7\linewidth]{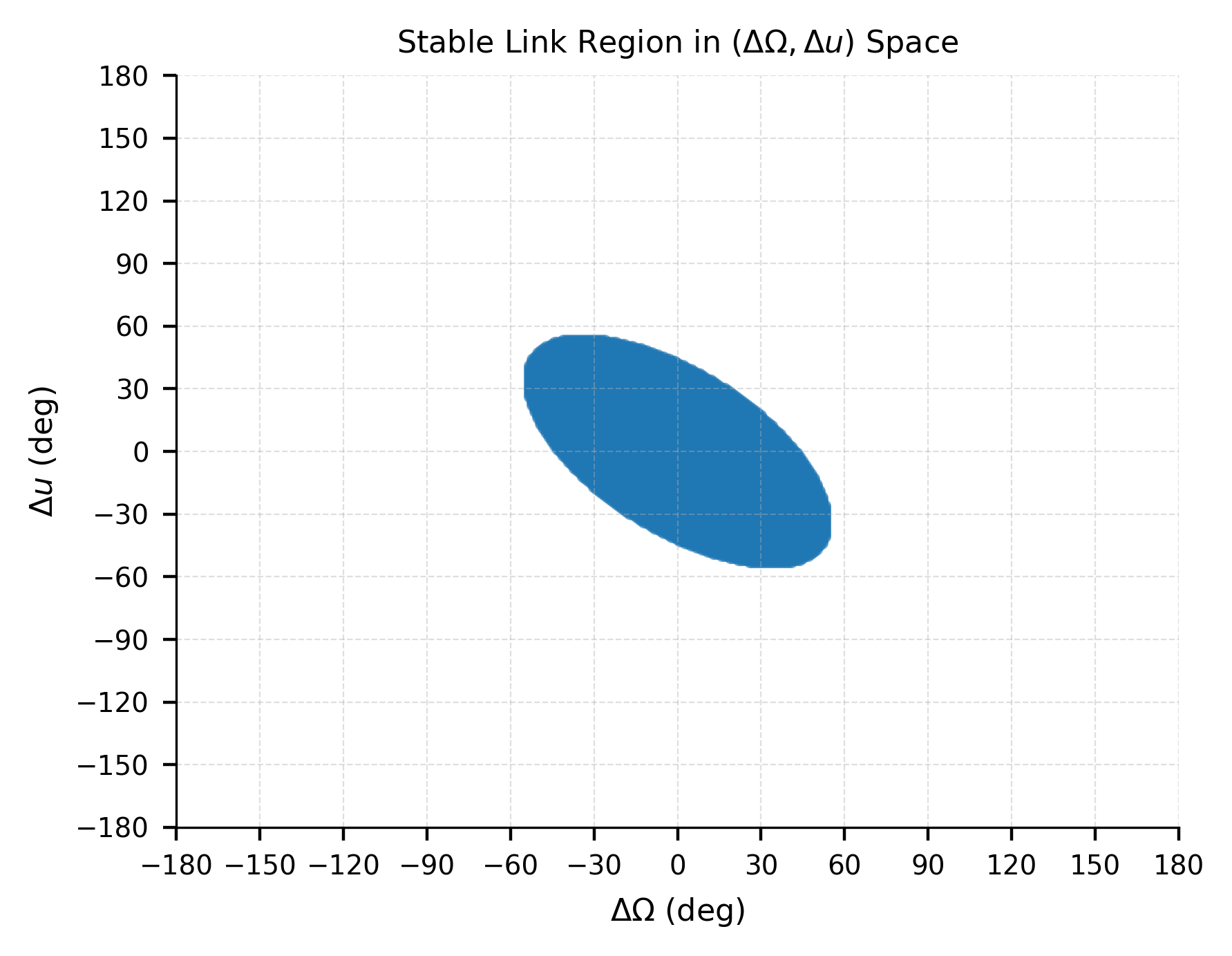}
\caption{Stable Links Region for Starlink Shell-1}

    \label{fig:stable-region}
\end{figure}

For each of the $360 \times 360$ combinations of angular offsets, we record the maximum distance encountered during the full orbital cycle. Finally, we keep the offsets whose maximum distance is within the stable-link threshold $D_{\mathrm{stab}}$. This procedure yields a stable-link region in $\Delta\Omega$--$\Delta u$ space that defines which pairs of satellites can maintain a stable link throughout the orbital cycle.
The results are shown in Figure~\ref{fig:stable-region} for the Starlink Shell-1.
Only links with offsets within the stable region are considered stable and will be candidates for later topology construction.

\begin{figure}[h]
    \centering
    \includegraphics[width=0.7\linewidth]{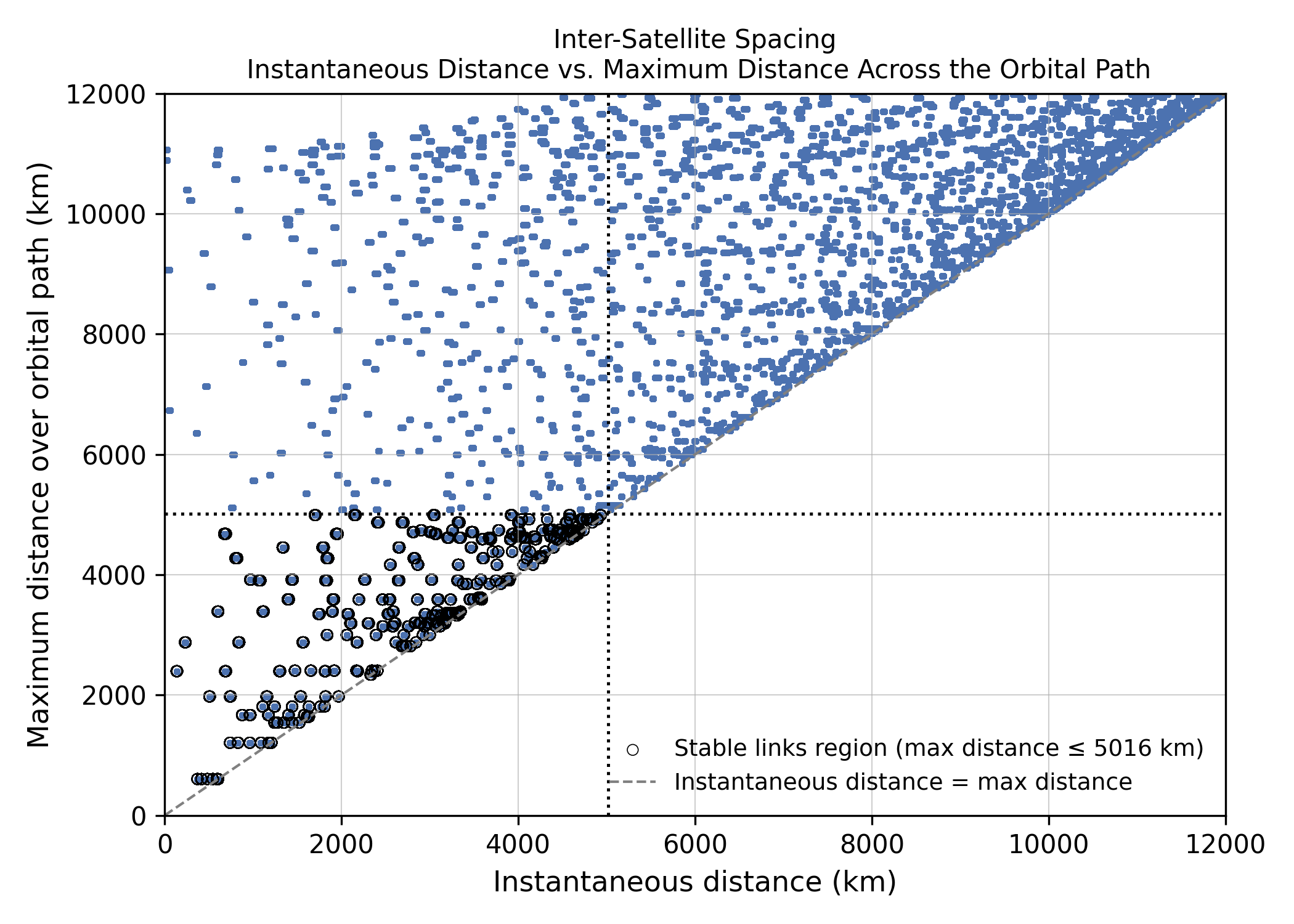}
    \caption{Satellite spacing in Starlink Shell-1. Instantaneous inter-satellite distance often underestimates maximum separation over the orbital path.}

    \label{fig:max_vs_instant}
\end{figure}

\textbf{Instantaneous vs. Worst-case Distance:}
One interesting observation from the stable-link computation is that a link's instantaneous distance, as measured in a single snapshot of the shell, often underestimates greatly the link's worst-case distance measured over the entire orbital period.
Figure~\ref{fig:max_vs_instant} compares instantaneous distance and maximum distance for all links in Starlink Shell-1. If a topology design relied on instantaneous distance, it would select many links that appear feasible at a given snapshot but actually have much larger maximum separation over the orbital cycle. This would lead to higher delay and potential link breakages if selected based on instantaneous distance.
Therefore, in our topology design, when we need to select or rank links based on distance, we use the worst-case distance over the orbital cycle rather than the instantaneous distance in order to ensure consistency and robustness over time.



\begin{figure*}[ht]
  \centering
  \makebox[\textwidth][c]{%
    \includegraphics[
      width=1.1\textwidth,
      trim=0 5.5cm 0 3.8cm,
      clip
    ]{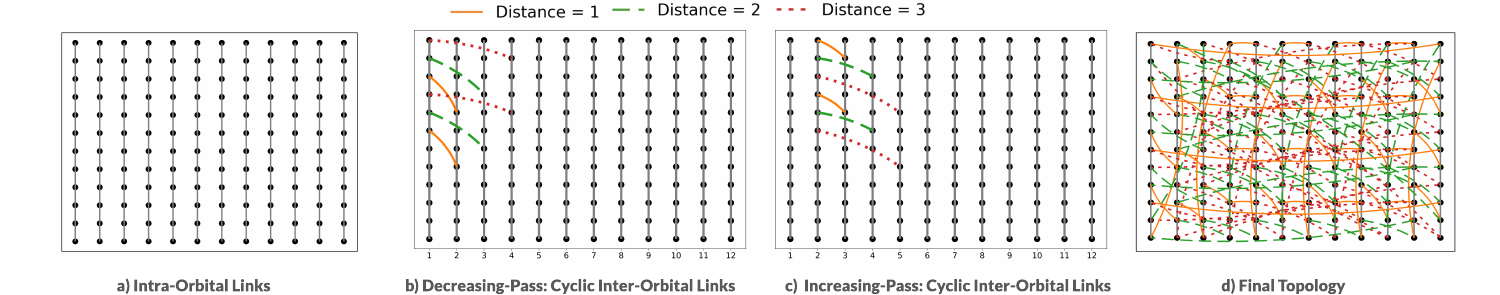}%
  }
  \caption{Construction of the Long--Short Links (LSL) topology on a toy $12\times 12$ satellite shell.}
  \label{fig:lsl_construction}
\end{figure*}

\subsection{Long-Short Links}
\label{subsec:lsl}


As explained in Section~\S\ref{sec:problems_topology_design}, a topology composed solely of short links, such as +Grid, cannot exploit long links across multiple orbits as shortcuts. A well-designed topology should have a balanced mix of short and long links to achieve low delay, low hop count, and high throughput. Motivated by this intuition, we designed the Long-Short Links (LSL) method to systematically select a blend of long and short links to achieve these goals without assuming full deployment or any topological symmetry.


LSL works in two stages. First, we add intra-orbital links to provide strong local connectivity, ensuring that once a packet reaches the correct orbit, it can efficiently reach the target satellite. Second, we incorporate a mix of short- to long-ranging inter-orbital links, giving the topology the ability to skip over multiple orbits when needed. This combination enables both fine-grained local movement and rapid long-range progression toward farther destinations. 

We describe the specific steps below using a toy $12\times 12$ constellation for illustration in Figure~\ref{fig:lsl_construction}.



\paragraph{\textbf{Intra-Orbit links}}
Within each orbit, we connect every satellite to its two immediate neighbors along the orbit, one ahead and one behind in orbital phase. This forms a ring in each plane and ensures strong local connectivity. This can be seen in Figure~\ref{fig:lsl_construction} (a). In case there are missing satellites due to partial deployment, we skip the missing one and connect to the next available satellite provided the link is stable. If the next available satellite is too far away, we will simply not connect to it.

\paragraph{\textbf{Inter-Orbit links.}}
This step adds a mix of short and long inter-orbital links. We define the inter-orbital separation between two satellites as the number of orbital planes separating them. An upper limit, \(D\), bounds how far links may span multiple orbits. 
For each orbit, satellites are sorted by angular phase (anomaly) and processed sequentially. For each satellite A, we add an inter-orbital link to another satellite B which is $d$ planes away. B is chosen as the closest feasible candidate in that plane, where feasibility is determined by the stable-link constraints, and the closeness is evaluated using the maximum link distance over the orbital cycle. As we traverse the satellites in the current plane, we vary $d$ in a repeating cycle of \(D, D\!-\!1, \ldots, 1\). That is, the first satellite in the plane connects to a candidate that is $D$ planes away, the second satellite connects to a candidate that is ($D\!-\!1$) planes away, and so on until we reach separation $1$, after which the cycle repeats. This process continues until all satellites in the current plane are processed or their degree limits are reached.

To balance the placement of long and short inter-plane links, we alternate the cycle direction across planes. If plane \(p\) follows the order \(D \rightarrow 1\), then plane \(p+1\) follows the reverse order \(1 \rightarrow D\). For example, in the \(12\times12\) toy constellation shown in Figure~\ref{fig:lsl_construction} (b) with \(D=3\), satellites in plane~1 are connected in a decreasing pass to the nearest feasible candidates in planes \(4,3,2\), while satellites in plane~2 follow an increasing pass and connect to planes \(3,4,5\) (Figure~\ref{fig:lsl_construction} (c)). This alternation distributes long and short inter-plane links more evenly across adjacent planes. We continue this process 
for all orbital planes. 
Note that both orbital phase and orbital-plane indices wrap around, so links connect circularly at the boundaries.

This cyclic process provides three key benefits:

\begin{itemize}
\item \textbf{Balanced link distribution.}  
The procedure spreads both long and short links uniformly across all orbital planes, preventing clusters of similar-length links and avoiding localized routing bottlenecks. Because each orbital plane receives links spanning all distances up to~$D$, the overall topology remains well balanced.

   \item \textbf{Delay-efficient long and short links.}  
    Long-range and short-range inter-orbital links are chosen based on geometric closeness, ensuring that even the longer shortcuts have relatively small physical distance. This keeps delay low while still enabling satellites to skip across multiple planes when needed to curb hop count.

  \item \textbf{Reduced path overlap and higher throughput.}
  By distributing inter-plane links across different separations and alternating the direction of the cycle across planes, the topology offers more diverse shortest paths. This reduces path overlap and link hot-spots, which in turn improves aggregate throughput (Section~\S\ref{subsec:traffic_capacity}).
\end{itemize}

Together, these effects produce a topology that maintains predominantly short, low-delay links while incorporating enough well-placed long inter-plane shortcuts to significantly reduce hop count. We demonstrate this in our evaluation in the next section (Section~\S\ref{subsec:path_efficiency}).

Even in this small $12\times12$ toy constellation (Figure~\ref{fig:lsl_construction}), LSL produces substantially shorter routes than +Grid. 
LSL achieves an average pairwise hop count of $4.23$, compared to $6.07$ under +Grid, corresponding to a hop-count reduction of approximately $30\%$. 
This structural improvement also translates into lower end-to-end delay, with LSL reducing average delay by about $12\%$.




\subsection{Simulated Annealing}


LSL is effective and efficient (as we will demonstrate in Section~\ref{sec:evaluation}) but intentionally structured. This structure imposes two limitations.

\begin{itemize}[noitemsep,topsep=2pt,leftmargin=*]
  \item \textbf{Restricted search space.} LSL constructs topologies by following a cyclic, pattern-based construction. While this makes the design fast and stable, it also limits exploration to a narrow family of graphs and may miss high-quality designs that do not conform to the assumed structure.
  \item \textbf{Assumes a structure.} LSL is particularly well matched to the common case of \textit{ISL limit} $=4$. It also assumes a shell configuration of many orbital planes.
\end{itemize}

To explore a broader range of topologies and do away with these limitations, we introduce a stochastic optimization method based on simulated annealing (SA). SA is a metaheuristic widely used for NP-hard combinatorial optimization problems, such as the Traveling Salesman Problem (TSP)~\cite{SAmcmc}. LEO topology design similarly involves selecting a feasible set of ISLs that optimizes multiple objectives under strict degree and feasibility constraints, resulting in a large discrete search space.



\subsubsection{Challenges in Using Simulated Annealing for Topology Design}
\label{subsec:sa_challenges}

SA requires evaluating an objective function after every proposed topology modification. For LEO constellations such as Starlink Shell-1, which consists of thousands of satellites, directly evaluating end-to-end path metrics at every step of the optimization is computationally prohibitive. 

Our ultimate topology design objectives are to minimize:
(i) average shortest-path delay between all satellite pairs,
(ii) average shortest-path hop count between all satellite pairs, and
(iii) link breakages over time.
However, evaluating delay or hop count requires computing all-pairs shortest paths using Dijkstra's algorithm, which takes \(O\!\left(|V|(|V|+|E|)\log|V|\right)\) time per evaluation. Embedding such computations inside SA would result in prohibitively long running times.
To quantify this cost, we benchmark repeated computations of the average shortest-path length (ASPL) using a graph matching the scale of Starlink Shell-1 (72 orbital planes $\times$ 22 satellites per plane). All experiments were conducted on modern commodity hardware (MacBook Pro with Apple M4 Pro). ASPL was computed using the standard implementation provided by the NetworkX library~\cite{networkX}, which implements Dijkstra's algorithm using binary heap.

\begin{table}[h]
\centering
\caption{Runtime cost of repeated ASPL computation on a Starlink Shell-1-scale topology.}
\label{tab:aspl_cost_shell1}
\begin{tabular}{l c}
\toprule
\textbf{Parameter} & \textbf{Value} \\
\midrule
Constellation            & Starlink Shell-1 \\
Number of satellites ($|V|$) & 1,584 \\
Number of links ($|E|$)      & 3,168 \\
ASPL runtime (mean)          & 2.09 seconds \\
ASPL runtime (min--max)      & 2.06--2.15 seconds \\
\midrule
Estimated time for 100k SA iterations & 58 hours \\
Estimated time for 200k SA iterations & 116 hours \\
\bottomrule
\end{tabular}
\end{table}

These results show that even for a single snapshot of Starlink Shell-1, recomputing ASPL at each SA iteration would require tens of hours for a single optimization run. As a result, directly optimizing shortest-path delay or hop count inside SA is infeasible in practice for mega-constellation-scale networks. To make SA tractable, we instead use constant-time ($O(1)$) surrogate objectives that can be updated incrementally after each local topology edit. These surrogates are chosen to correlate strongly with delay and hop count, enabling efficient exploration of the topology design space without repeatedly recomputing all-pairs shortest paths.

\subsubsection{Surrogate objectives.}
During SA optimization, after each proposed topology edit, we compute and optimize three constant-time surrogate objectives:
\begin{enumerate}[noitemsep]
    \item \emph{Mean link length} (\(L\)), to be minimized, since shorter links generally reduce delay.
    \item \emph{Fraction of long links} (\(M\)), to be maximized; long-range links act as shortcuts that lower hop count.
    \item \emph{Edge utilization} (\(U\)), to be maximized, since using more of the available degree budget increases path diversity and resilience.
\end{enumerate}

By encouraging shorter average link lengths while preserving a useful fraction of long-range shortcuts, and by promoting high edge utilization, these surrogates jointly improve delay, hop count, and link density. We further validate the effectiveness of these surrogate objectives in our evaluation (Section~\S\ref{subsec:path_efficiency}).

Moreover, because all candidate edges are drawn from the stable-link set, the resulting topologies naturally reduce link breakages over multi-day horizons (Section~\S\ref{subsec:link_breakages}).


\subsubsection{Approach Overview}
Concretely, we start from an initial random feasible topology built only from the stable-link set. The search then proceeds iteratively: at each step, we perform a \textit{proposal} by adding a randomly chosen link from stable-link set; if this would exceed the degree budget at either endpoint, we instead perform a swap by removing a conflicting incident edge to restore feasibility. We evaluate the proposal using the weighted surrogate score, accept it if it improves the score, and otherwise accept or reject it probabilistically under the simulated annealing rule. This process repeats until the iteration budget is reached.

\subsubsection{Implementation.}
We sort the orbits based on their RAAN values and define the inter-plane distance as
\[
\Delta_{\text{planes}}(i,j)=\min\bigl\{|i-j|,\;N-|i-j|\bigr\},
\]
where \(i\) and \(j\) are the plane indices of two satellites and \(N\) is the total number of planes.

Links with $\Delta_{\text{planes}} \le 3$ are \emph{neighbors across planes}; otherwise they are \emph{non-neighbors} (long-range). Let $d_{uv}$ denote the geometric length of edge $(u,v)$, measured as its maximum separation over an orbital cycle. We normalize link lengths by the maximum stable-link distance $L_{\max}=D_{\mathrm{stab}}$ (in km). With a per-satellite degree limit $deg_{\max}$, the maximum feasible edge budget is
\[
E_{\max}=\left\lfloor \frac{deg_{\max}\,|V|}{2}\right\rfloor .
\]
We maintain the following normalized surrogates:

\[
\text{L}=\frac{\sum_{(u,v)\in E} d_{uv}}{|E|\,L_{\max}}\times 100,\quad
\text{M}=\frac{|E_{\text{long\_range}}|}{|E|}\times 100,\quad
\text{U}=\frac{|E|}{E_{\max}}\times 100.
\]

\begin{quote}
\textbf{Algorithm: Simulated Annealing for ISL Topology (High Level)}

\textit{Inputs.} A set of stable links; a max degree per satellite; weights $(\alpha_L,\alpha_U,\alpha_M,)$; a temperature schedule; an iteration limit.

\begin{enumerate}[noitemsep]

\item \textbf{Initialization.} Build a minimally connected graph from stable-link set.

\item \textbf{Proposal (edit).} Pick a stable link uniformly at random and propose adding it. If both endpoints stay within the degree limit, tentatively add it. Otherwise, swap by adding the link and randomly removing one incident edge at each violated endpoint to restore feasibility. Proceed only if the topology remains connected. Otherwise, discard the edit and continue to next iteration.

\item \textbf{Scoring and acceptance.} Recompute the surrogate metrics on the tentative graph, including the mean link length $L$ (smaller is better), the fraction of long-range edges $M$ (larger is better), and the edge utilization $U$ relative to the allowed maximum (larger is better).
  We form a weighted improvement score:
  \[
    \Delta \;=\; \alpha_L\,(L_{\text{old}}-L_{\text{new}})
                  \;+\; \alpha_M\,(M_{\text{new}}-M_{\text{old}})
                  \;+\; \alpha_U\,(U_{\text{new}}-U_{\text{old}}).
  \]
  We accept the tentative topology according to the standard simulated-annealing rule:
  \[
    \text{accept} \;=\;
      \begin{cases}
        \text{yes}, & \Delta \ge 0,\\[2pt]
        \text{yes with prob. } \exp(\Delta/T), & \Delta < 0.\\
      \end{cases}
  \]

\item \textbf{Cooling and stopping.} Decrease the temperature after each step, e.g.,
  \[
    T \leftarrow \max\!\big(T_{\min},\, \rho\,T\big)\quad\text{with } \rho\in(0,1),
  \]
  and stop after reaching the step limit.

\item \textbf{Add residual Links (fill remaining degree).} after reaching the step limit, some satellites may still have unused degree. We then greedily add additional stable links in order of increasing geometric distance (shortest first), subject to the degree budget. This final step completes any remaining feasible connections and can further reduce delay and hop count.

\end{enumerate}

\end{quote}

\paragraph{Tunability}
By adjusting the surrogate objective weights, the network provider can tune SA to generate a \emph{family} of candidate topologies, spanning different trade-offs between hop count and delay, as we demonstrate in Section~\S\ref{subsubsec:synthetic_compare}.

\subsection{Incremental Updates}
\label{subsec:incremental_updates}

Real deployments evolve over time: satellites are added, de-orbited, or temporarily unavailable, and link feasibility changes as well. Recomputing a topology from scratch every day can therefore introduce unnecessary link changes and cause network instability. Instead, we update the topology at a daily cadence by preserving as much of previous day's design as possible and changing only what is needed to remain feasible.

On day~1, we compute the topology from scratch. On each subsequent day, we rebuild the day's stable-link set and update the previous topology following the steps below:

\paragraph{\textbf{Incremental Updates for LSL.}}
\begin{enumerate}[noitemsep,topsep=0pt]
  \item \textbf{Retain feasible links.} Retain all links from day $t$ whose endpoints remain active on day $t{+}1$ and whose links are still feasible under the new stable-link set, without exceeding the degree budget.
  \item \textbf{Refresh intra-plane links.} Recompute LSL’s intra-plane links for the current day to maintain local connectivity after node additions/removals.
  \item \textbf{Repair connectivity.} If the retained graph becomes disconnected, iteratively add feasible links that connect smaller components to the largest component until the topology is connected. 
  \item \textbf{LSL application.} With connectivity restored, add additional inter-plane links using LSL’s standard link-selection procedure over the remaining stable-link candidates, until satellites reach the degree budget.
\end{enumerate}

\paragraph{\textbf{Incremental Updates for SA.}}
\begin{enumerate}[noitemsep,topsep=0pt]
  \item \textbf{Retain feasible links.} Retain all feasible links from the previous day that respect the current degree budget.
  \item \textbf{Repair connectivity.} If the retained graph becomes disconnected, iteratively add feasible links from stable-link set that connect smaller components to the largest component until the topology is connected. 
  \item \textbf{Re-optimize minimally.} Starting from this warm-start topology, run SA restricted to the current day’s stable-link candidates, making only the edits necessary to improve the surrogate objectives under the degree budget.
\end{enumerate}

We demonstrate the efficacy of our incremental update procedures in Section~\S\ref{subsec:robustness}.
\section{Evaluation}
\label{sec:evaluation}

We evaluate our topology designs using a comprehensive methodology that considers both controlled, fully deployed shells and real Starlink deployments.



\subsection{Datasets}
\label{subsec:datasets}
\textbf{{Synthetic shells (controlled, fully deployed):}}
We generate idealized, fully populated shells from synthesized two-line element (TLE) data to enable controlled comparisons without deployment effects. Specifically, we construct (i) a synthetic Starlink Shell-1 instance with $72\times 22$ satellites at 550\,km altitude and $53^\circ$ inclination, and (ii) a synthetic Amazon Kuiper instance with $34\times 34$ satellites at 630\,km altitude and $51.9^\circ$ inclination. We select Starlink Shell-1 because it represents a dense LEO configuration, and the Amazon Kuiper shell because it provides a contrasting, sparser configuration at a different altitude and inclination.

\textbf{Real Starlink deployment data:}
We use daily Starlink Shell-1 TLE data spanning October--December 2024 (90~days), collected from Space-Track~\cite{spacetrack_usspacecom}. Space-Track provides one TLE per satellite per day, which we use as the daily snapshot. TLEs include orbital elements (e.g., RAAN and inclination), which we use to identify orbital planes and estimate each satellite's along-orbit position.  We then run \texttt{sgp4} to propagate each TLE and obtain per-day ECI 3D positions~\cite{python-sgp4}. The resulting snapshots capture real operational dynamics over a long period of 3-months, including satellite turnover, orbital maintenance, and day-to-day variability in feasible links.





\subsection{Baselines}
\label{subsec:baselines}

We construct our designs under two ISL degree limits: 4-ISL (e.g., a satellite can connect to up to 4 others), to enable comparison with prior work, and 3-ISL, which reflects current optical terminal capabilities of Starlink~\cite{starlink_technology}.  We compare against three standard baselines and a theoretical reference:
\begin{itemize}[noitemsep,topsep=0pt]
  \item \textbf{+Grid}~\cite{delyHandley,IPCGrid}: connects each satellite to two intra-plane neighbors and nearest satellites in adjacent planes.
  \item \textbf{3-ISL-Grid}~\cite{3_isl_grid}: each satellite maintains two intra-plane links and a single inter-plane link. Inter-plane links alternate between the left and right adjacent planes (brick-wall style).

  \item \textbf{Motif}~\cite{motif}: applies a repeating two-link pattern (motif) across the entire shell. 
  \item \textbf{Theoretical floor:} In this topology, a satellite can form links with any number of satellites if they are in feasible optical range (ISL degree constraint removed). This graph is infeasible in practice but quantifies how close a constrained design approaches the theoretical floor.
\end{itemize}

Note that Motif requires an explicit 4-ISL configuration and does not operate under a 3-ISL constraint. The +Grid baseline requires a 4-ISL constraint, while the 3-ISL Grid baseline applies under the 3-ISL setting.

\vspace{-0.2cm}

\subsection{Metrics}
\label{subsec:metrics}

We evaluate our designs along three attributes: path quality, network capacity, and robustness. For each attribute, we compute the following metrics.

{\textbf{Path quality:}}
We evaluate end-to-end path quality over all source--destination satellite pairs using the \emph{average shortest-path delay}, defined as the mean propagation delay along shortest-distance paths, and the \emph{average shortest-path hop count}, defined as the mean number of hops along those paths. We also report the empirical distributions of shortest-path delay and hop count.

{\textbf{Network capacity:}}
We estimate network capacity by measuring aggregate throughput under load. Throughput is computed using max--min fair bandwidth allocation under a traffic matrix with single shortest-path routing, and reported as the total delivered throughput across all flows.

{\textbf{Robustness:}}
We quantify robustness under real operational dynamics by measuring link breakages in real data, defined as the fraction of links selected on day~$t$ that become infeasible on day~$t{+}1$. We additionally conduct an ablation study under controlled turnover using a Starlink Shell-1 configuration, measuring changes in delay and hop count when the topology is updated incrementally rather than recomputed from scratch, under shell growth and shrinkage scenarios.

We first report results on fully deployed synthetic constellations (Starlink Shell-1 and an Amazon Kuiper shell) to enable controlled comparisons across topology designs. We then present results on real, operational Starlink Shell-1 deployments.

\begin{table}[ht]
\centering
\begin{adjustbox}{max width=\linewidth}
\begin{tabular}{lccc}
\toprule
& \multicolumn{3}{c}{\textbf{Starlink Shell-1 (72$\times$22)}} \\
\cmidrule(lr){2-4}
\textbf{Method} 
& \textbf{Delay (ms)} 
& \textbf{\shortstack{Delay Stretch to\\ Theoretical Minimum ($\times$)}} 
& \textbf{Hop Count} \\
\midrule
Motif (low delay)   & 48.4 & 1.34 & 23.3 \\
Motif (balanced)    & 48.9 & 1.36 & 21.5 \\
+Grid               & 60.6 & 1.68 & 23.5 \\
Long--Short Links   & \textbf{46.9} & \textbf{1.30} & 8.8 \\
SA (low delay)      & 49.0 & 1.36 & 12.5 \\
SA (low hop count)  & 61.2 & 1.70 & \textbf{8.5} \\
SA (balanced)       & 52.0 & 1.44 & 10.6 \\
\midrule
& \multicolumn{3}{c}{\textbf{Amazon Kuiper (34$\times$34)}} \\
\cmidrule(lr){2-4}
\textbf{Method} 
& \textbf{Delay (ms)} 
& \textbf{\shortstack{Delay Stretch to\\ Theoretical Minimum ($\times$)}} 
& \textbf{Hop Count} \\
\midrule
Motif (low delay)   & \textbf{44.4} & \textbf{1.22} & 19.0 \\
Motif (balanced)    & 48.9 & 1.34 & 17.0 \\
+Grid               & 56.3 & 1.55 & 17.1 \\
Long--Short Links   & \textbf{44.4} & \textbf{1.22} & 9.8 \\
SA (low delay)      & 49.3 & 1.35 & 10.9 \\
SA (low hop count)  & 62.8 & 1.73 & \textbf{8.5} \\
SA (balanced)       & 53.9 & 1.48 & 9.4 \\
\bottomrule
\end{tabular}
\end{adjustbox}
\caption{Performance on two synthetic shells with ISL limit $=4$. \textbf{Starlink Shell-1:} $72\times22$ at 550\,km and $53^\circ$. \textbf{Amazon Kuiper:} $34\times34$ at 630\,km and $51.9^\circ$. The theoretical delay floors are 36.1\,ms (Starlink Shell-1) and 36.4\,ms (Kuiper); we compute stretch relative to these floors.}
\label{tab:shell1_kuiper_perf}
\end{table}


\begin{figure*}[ht]
  \centering
  \captionsetup[subfigure]{font=scriptsize}

  \begin{subfigure}{0.24\textwidth}
    \centering
    \includegraphics[width=\linewidth]{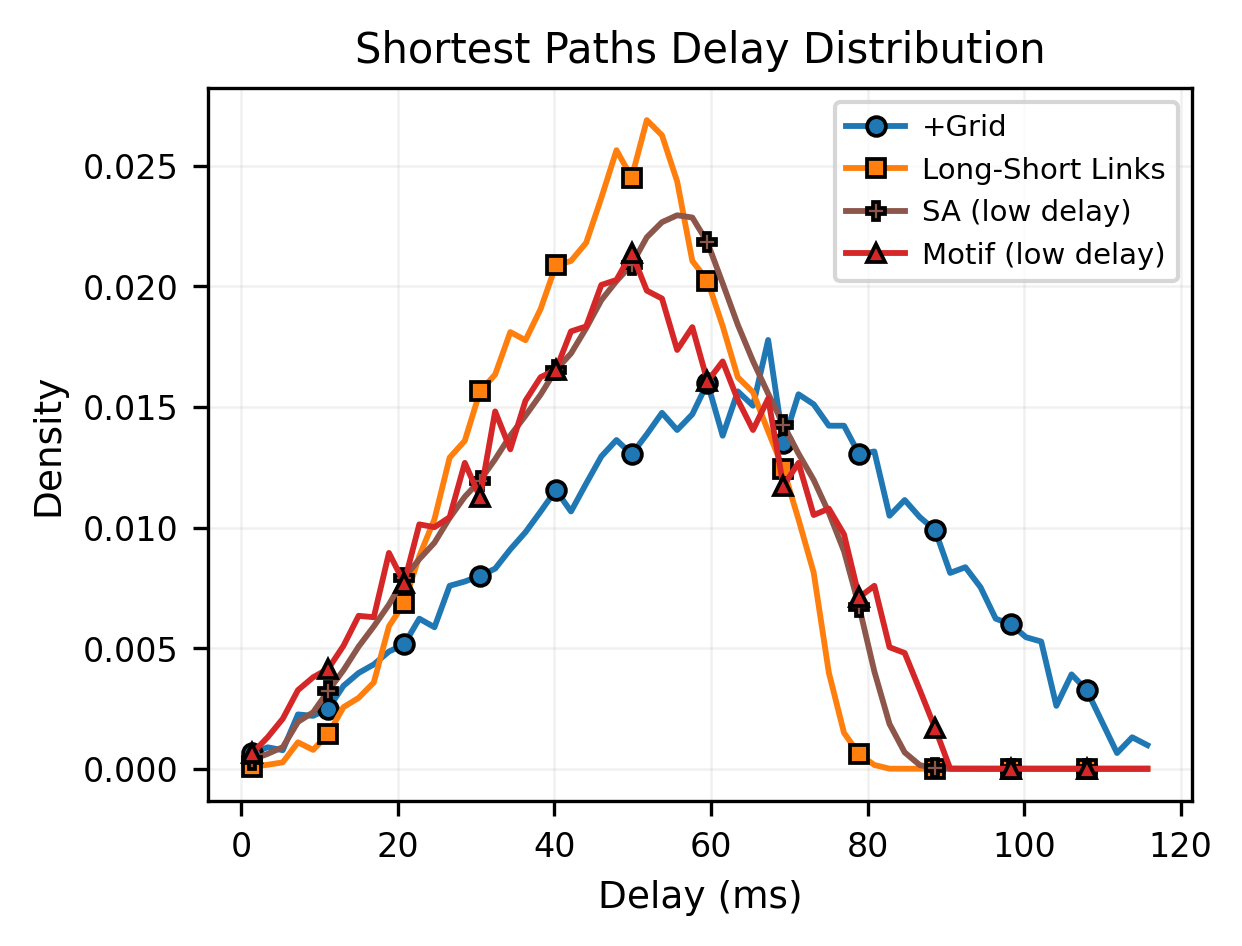}
    \caption{Delay distribution}
    \label{fig:syn_shell1_delay_dist}
  \end{subfigure}\hfill
  \begin{subfigure}{0.24\textwidth}
    \centering
    \includegraphics[width=\linewidth]{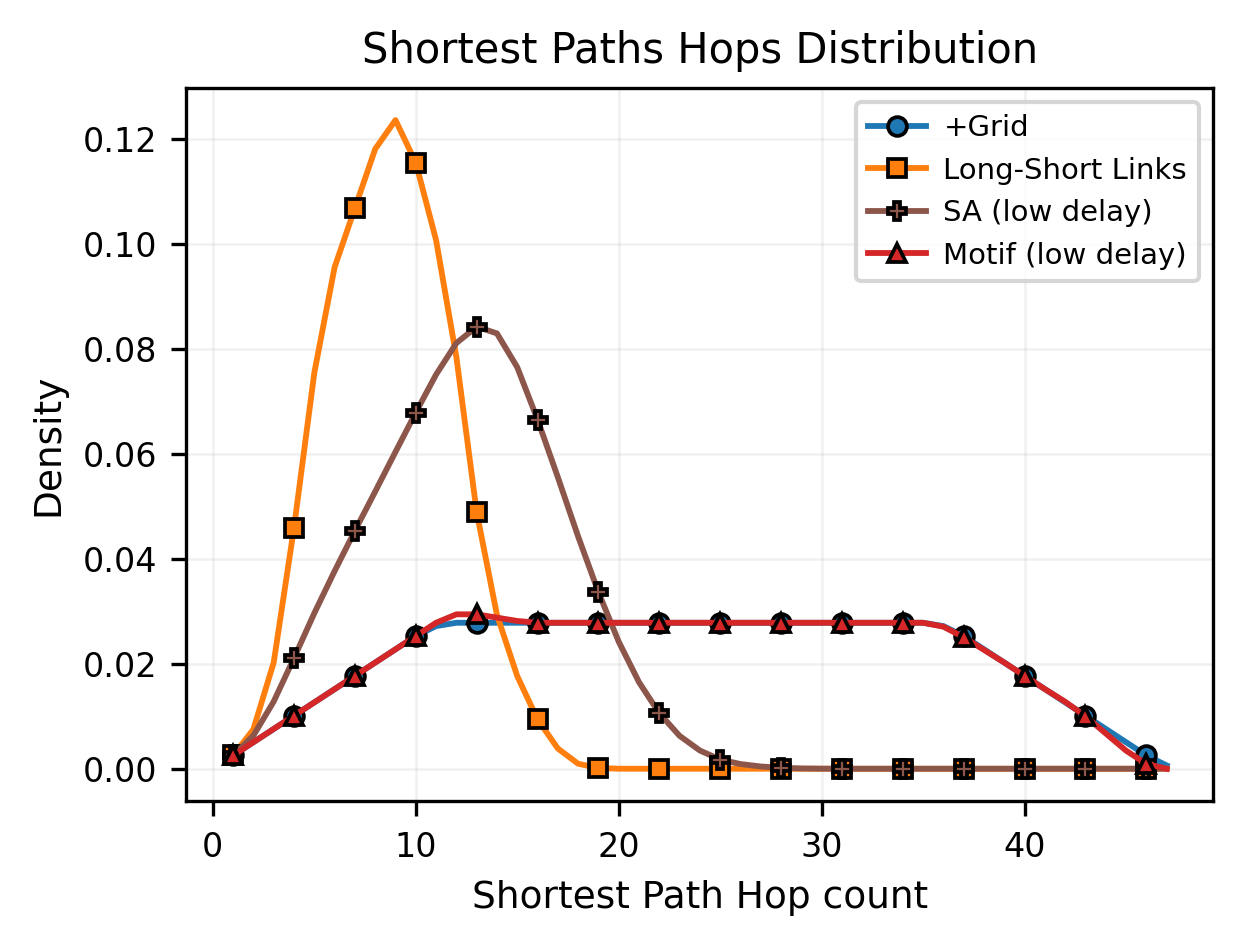}
    \caption{Hop Count distribution}
    \label{fig:syn_shell1_hop_dist}
  \end{subfigure}\hfill
  \begin{subfigure}{0.24\textwidth}
    \centering
    \includegraphics[width=\linewidth]{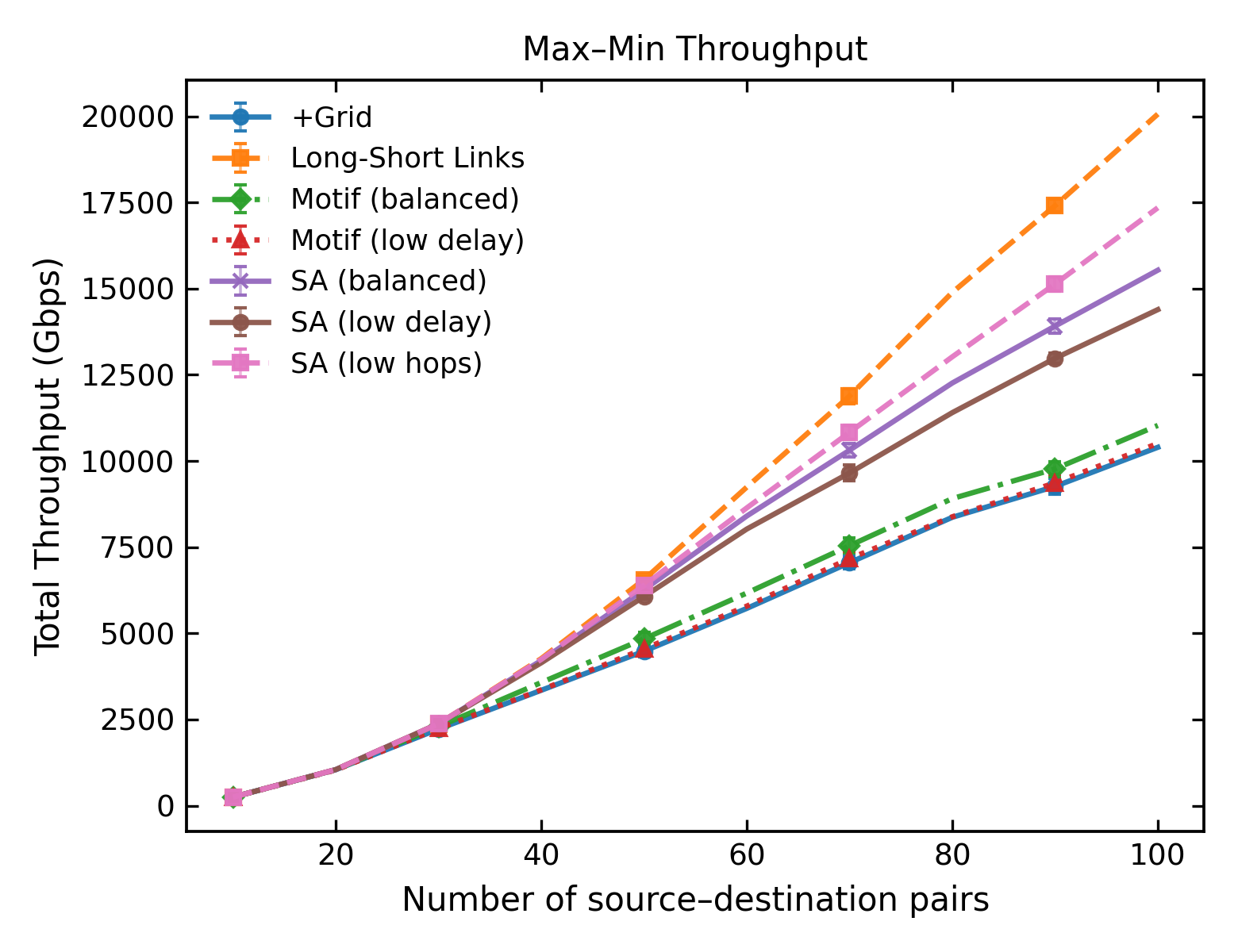}
    \caption{Max-Min Fairness Throughput}
    \label{fig:throughput_shell1_syn}
  \end{subfigure}\hfill
  \begin{subfigure}{0.24\textwidth}
    \centering
    \includegraphics[width=\linewidth]{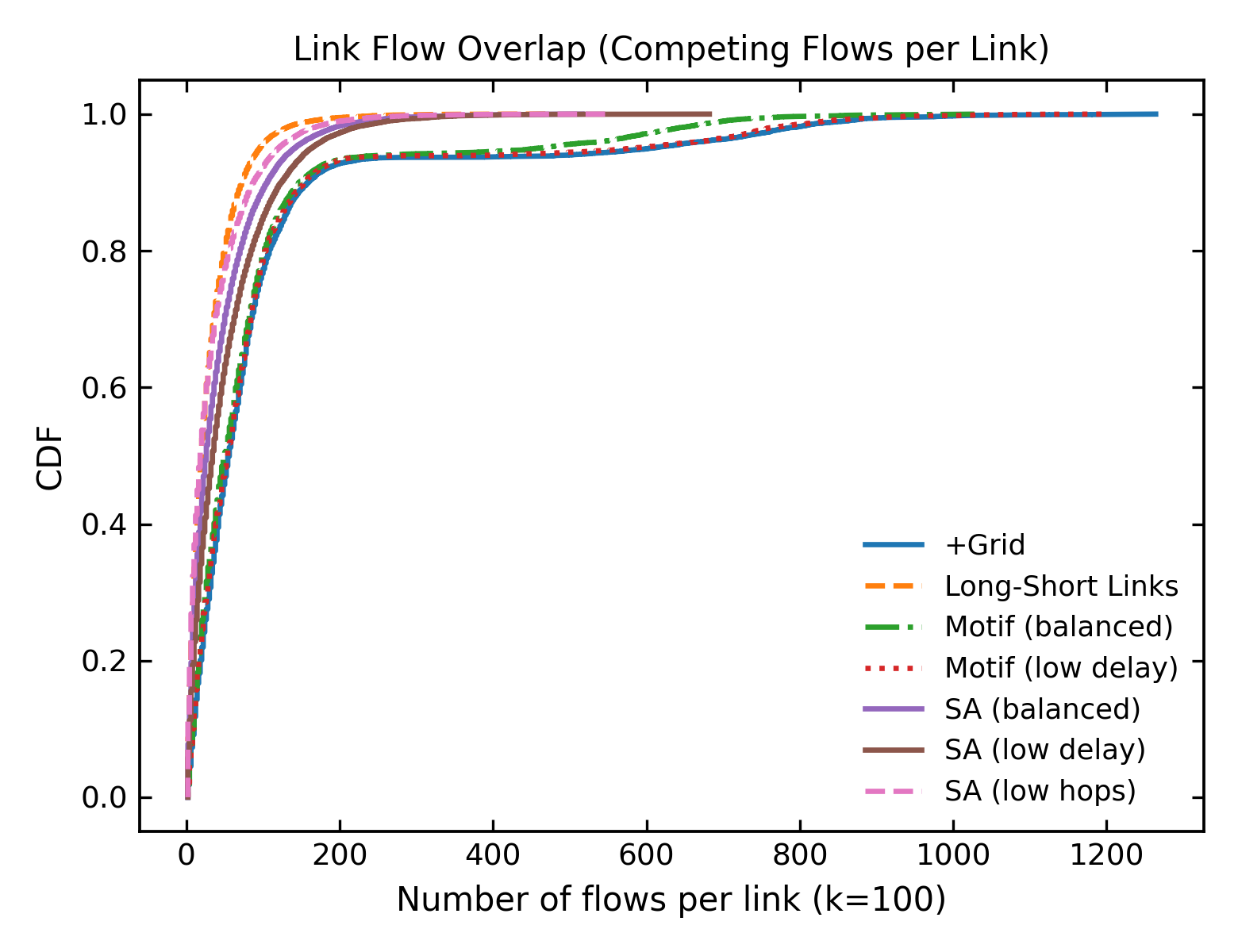}
    \caption{Flows per link for 100 src-dest pairs.}
    \label{fig:flows_per_link_syn}
  \end{subfigure}

  \vspace{0.35em}

  \begin{subfigure}{0.25\textwidth}
    \centering
    \includegraphics[width=\linewidth]{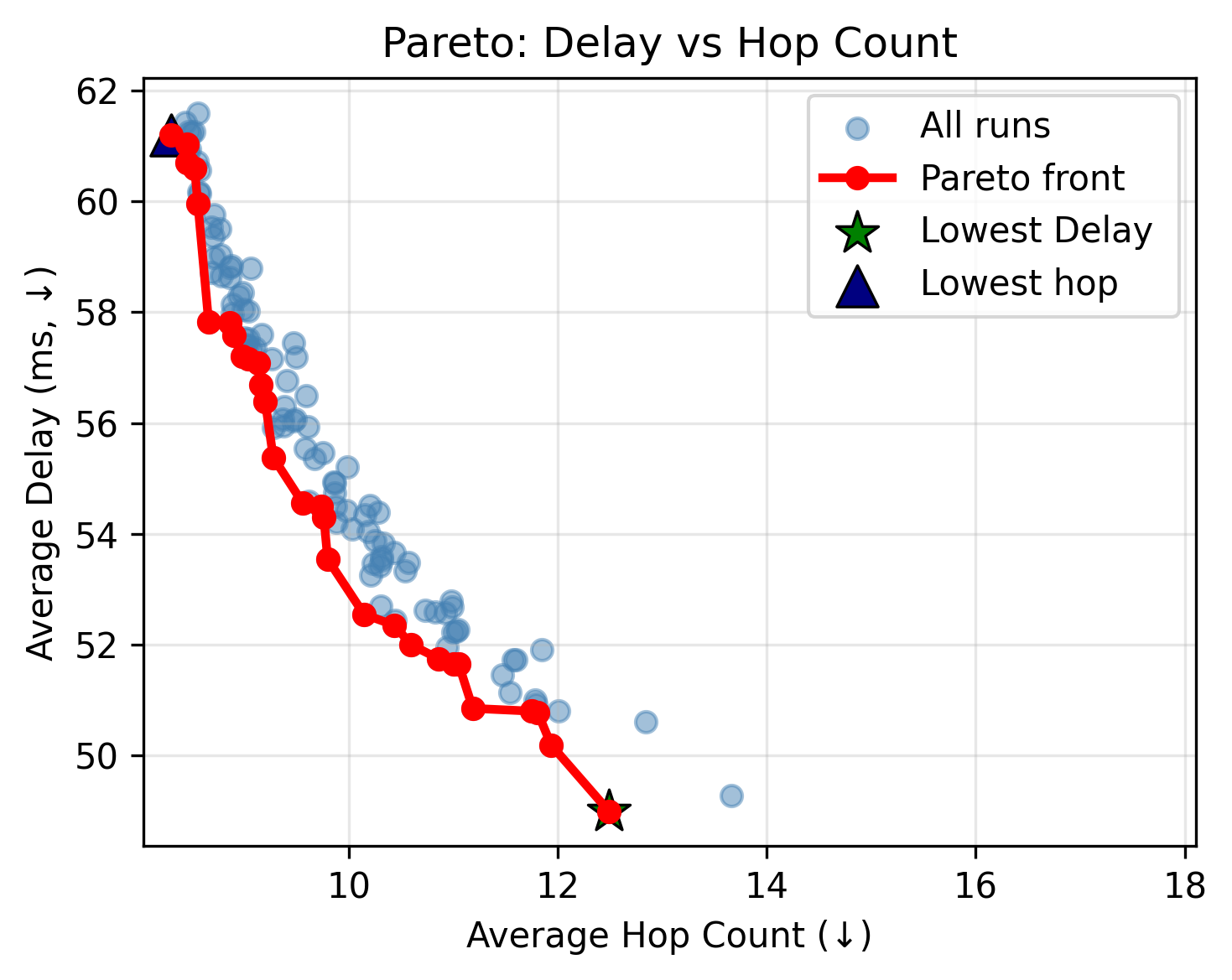}
    \caption{Pareto frontier over SA designs.}
    \label{fig:pareto_all}
  \end{subfigure}\hfill
  \begin{subfigure}{0.25\textwidth}
    \centering
    \includegraphics[width=\linewidth]{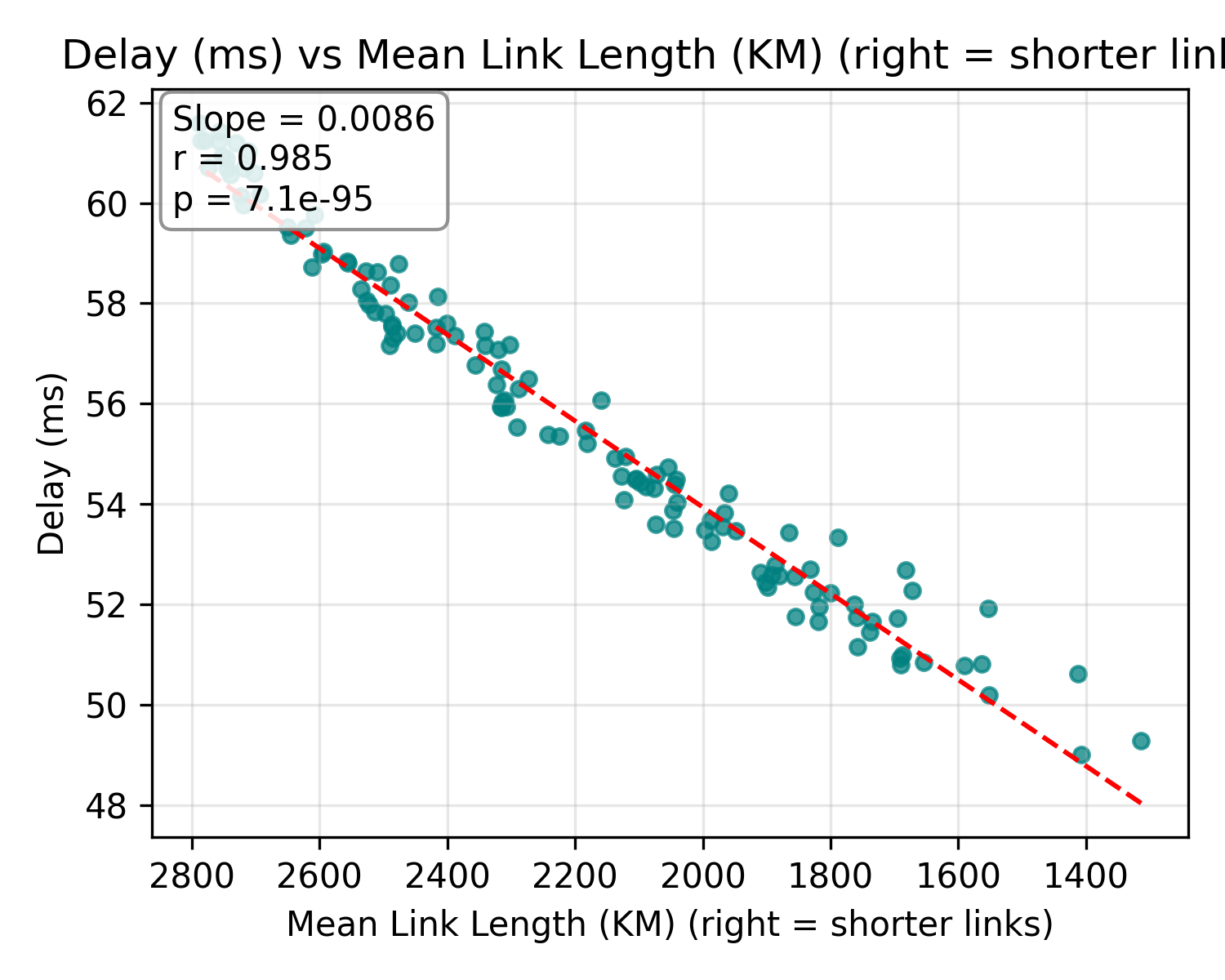}
    \caption{Delay vs.\ mean link length $L$ under SA designs.}
    \label{fig:len_vs_delay_corr_sub}
  \end{subfigure}\hfill
  \begin{subfigure}{0.25\textwidth}
    \centering
    \includegraphics[width=\linewidth]{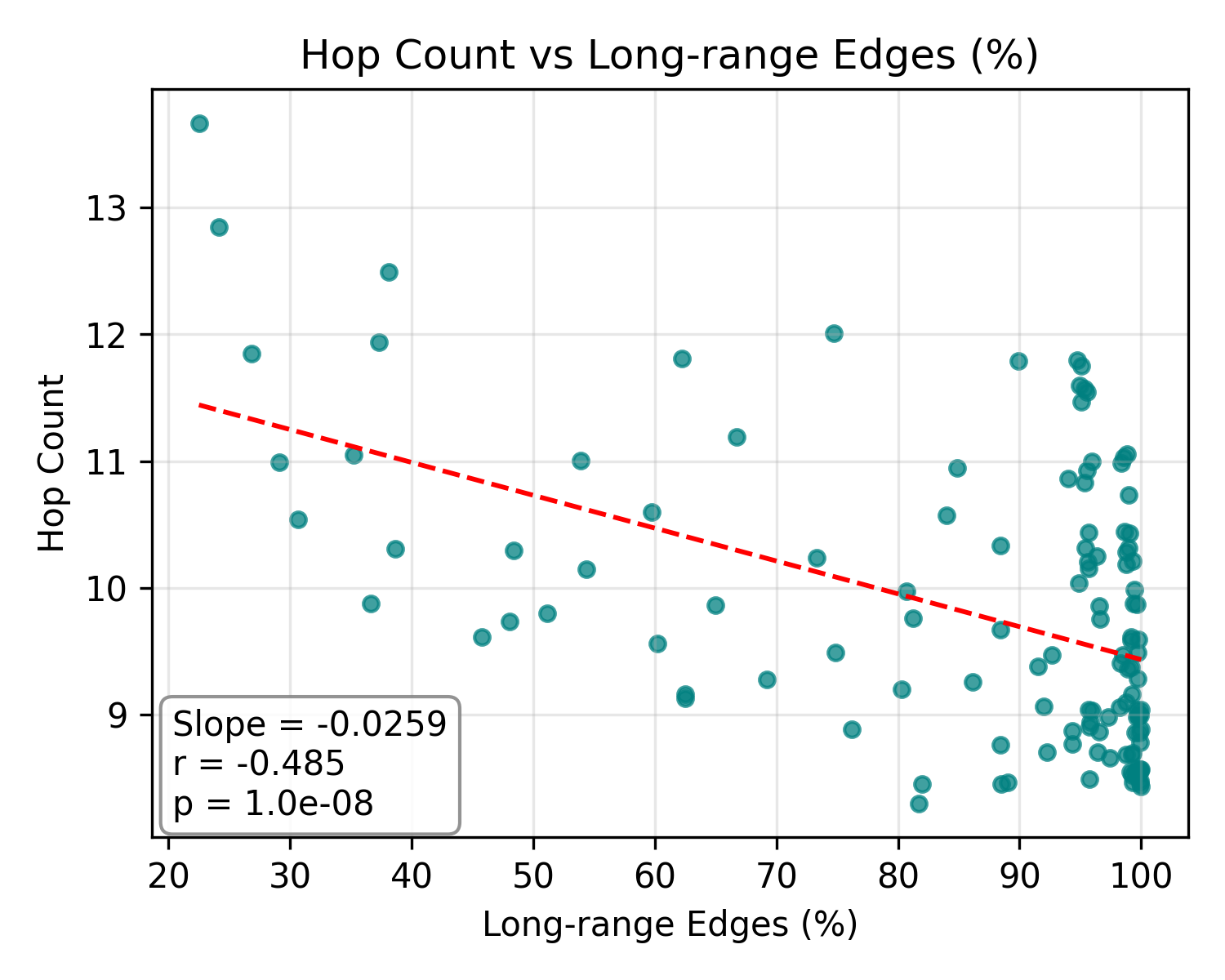}
    \caption{Hops vs.\ long-range fraction $M$ under SA designs.}
    \label{fig:long_vs_hop_corr_sub}
  \end{subfigure}
  \caption{Synthetic Starlink Shell-1 evaluation (ISL Limit=4).}
  \label{fig:synthetic_all_at_once}
\end{figure*}

\subsection{Synthetic Constellation Results}
\label{subsubsec:synthetic_compare}

\textbf{Representative designs:} 
SA is a tunable framework that can produce different topology designs by adjusting the surrogate objective weights. For each shell (Starlink Shell-1 and the Amazon Kuiper shell), we report three SA designs: a \emph{low-delay} design, a \emph{low-hop} design, and a \emph{balanced} design that trades off hop count and delay. Similarly, for Motif we report two designs (\emph{low-delay} and \emph{balanced}), since its low-hop solution is nearly identical to the balanced solution. The corresponding parameters for our approaches are listed in Appendix Table~\ref{tab:param_settings}.

\subsubsection{\textbf{Path Quality}}
\label{subsec:path_efficiency}


Table~\ref{tab:shell1_kuiper_perf} compares our methods against +Grid and Motif on synthetic Shell-1 and the Amazon Kuiper shell under 4-ISL configuration. Table~\ref{table:3-isl-two-shells} additionally reports results under a stricter 3-ISL configuration.
 

{\textbf{Findings:}}
For 4-ISL setups (Table~\ref{tab:shell1_kuiper_perf}) in Starlink Shell-1, both LSL and SA outperform +Grid and Motif in hop count by a large margin. 
On Shell-1, LSL reduces average hop count from about \(21\)–\(23\) (Motif/\(+\)Grid) to \(8.8\), approximately \(59\%\)–\(63\%\) reduction. LSL also achieves the lowest delay (46.9 ms) of all approaches and is within $1.3\times$ of the theoretical floor.

SA offers a broader operating range. The \emph{low-delay} SA design nearly matches Motif's delay while reducing hop count by roughly half. The \emph{low-hop} SA design achieves the lowest hop count of all approaches (8.5), corresponding to a 60\% reduction relative to Motif (\emph{balanced}) and a 64\% reduction relative to +Grid.

The same pattern holds on Amazon Kuiper: LSL and SA reduce hop count well below the baselines, with LSL offering the strongest delay--hop trade-off. 
Importantly, these gains hold across the full path distributions, not only in averages, as seen in Figure~\ref{fig:syn_shell1_delay_dist} and Figure~\ref{fig:syn_shell1_hop_dist}.

\begin{table}[h]
\centering
\resizebox{\linewidth}{!}{%
\begin{tabular}{lcc|cc}
\toprule
& \multicolumn{2}{c|}{\textbf{Starlink Shell-1 (72$\times$22)}} 
& \multicolumn{2}{c}{\textbf{Amazon Kuiper (34$\times$34)}} \\
\cmidrule(lr){2-3}\cmidrule(lr){4-5}
\textbf{Method}
& \textbf{Delay (ms)}
& \textbf{Hop Count}
& \textbf{Delay (ms)}
& \textbf{Hop Count} \\
\midrule
3-ISL-Grid               & 146.4 & 36.6  & 76.0 & 19.8 \\
Long--Short Links   & 80.0 & 13.3  & 65.7 & 14.6 \\
SA (low delay)      & \textbf{61.6} & 15.9  & \textbf{59.8} & 13.4 \\
SA (low hop count)  & 80.6 & \textbf{11.5} & 90.0 & \textbf{11.3} \\
SA (balanced)       & 66.5 & 13.4  & 70.9 & 12.2 \\
\bottomrule
\end{tabular}%
}
\caption{Performance on synthetic shells with ISL limit$=3$.}
\label{table:3-isl-two-shells}
\end{table}

For the 3-ISL setting (Table~\ref{table:3-isl-two-shells}), our methods significantly outperform the 3-ISL-Grid baseline. On Starlink Shell-1, SA (low delay) reduces average delay from \(146.4\) (3-ISL-Grid) to \(61.6\) (\(58\%\)), while SA (low hop) reduces average hop count from \(36.6\) to \(11.5\) (\(69\%\)). LSL also reduces delay from \(146.4\) to \( 80\)~ms (\( 45\%\)) and reduces hop count from \(36.6\) to \(13.3\) (\( 64\%\)).  On Amazon Kuiper shell with 3-ISL configuration, LSL and SA reduce hop count and delay well below the baselines, with SA offering the best delay--hop trade-off.



\begin{figure*}[ht]
  \centering
  \captionsetup[subfigure]{font=scriptsize}

  \begin{subfigure}{0.245\textwidth}
    \centering
    \includegraphics[width=\linewidth]{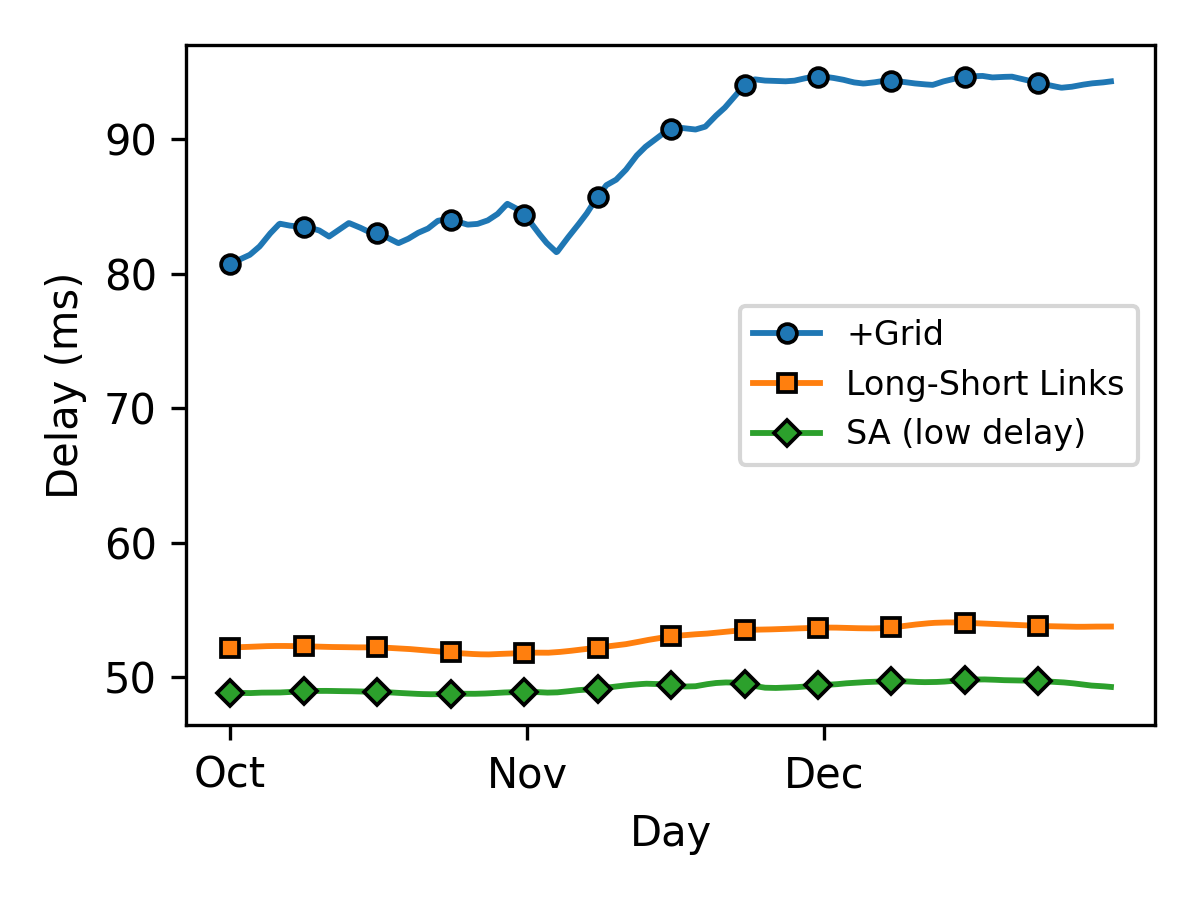}
    \caption{Delay over time (ISL=4).}
    \label{fig:real_delay_over_time_4}
  \end{subfigure}\hfill
  \begin{subfigure}{0.245\textwidth}
    \centering
    \includegraphics[width=\linewidth]{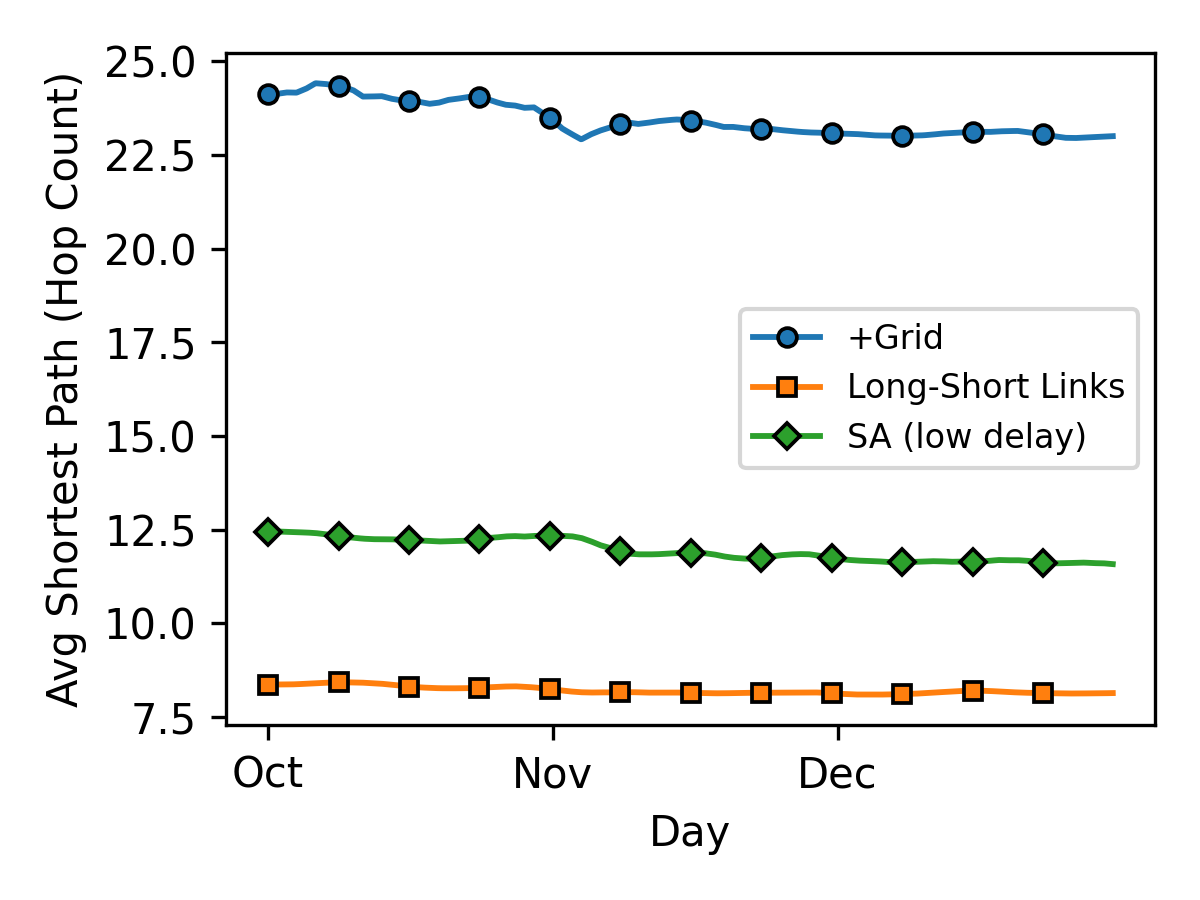}
    \caption{Hop count over time (ISL=4).}
    \label{fig:real_hop_over_time_4}
  \end{subfigure}\hfill
  \begin{subfigure}{0.245\textwidth}
    \centering
    \includegraphics[width=\linewidth]{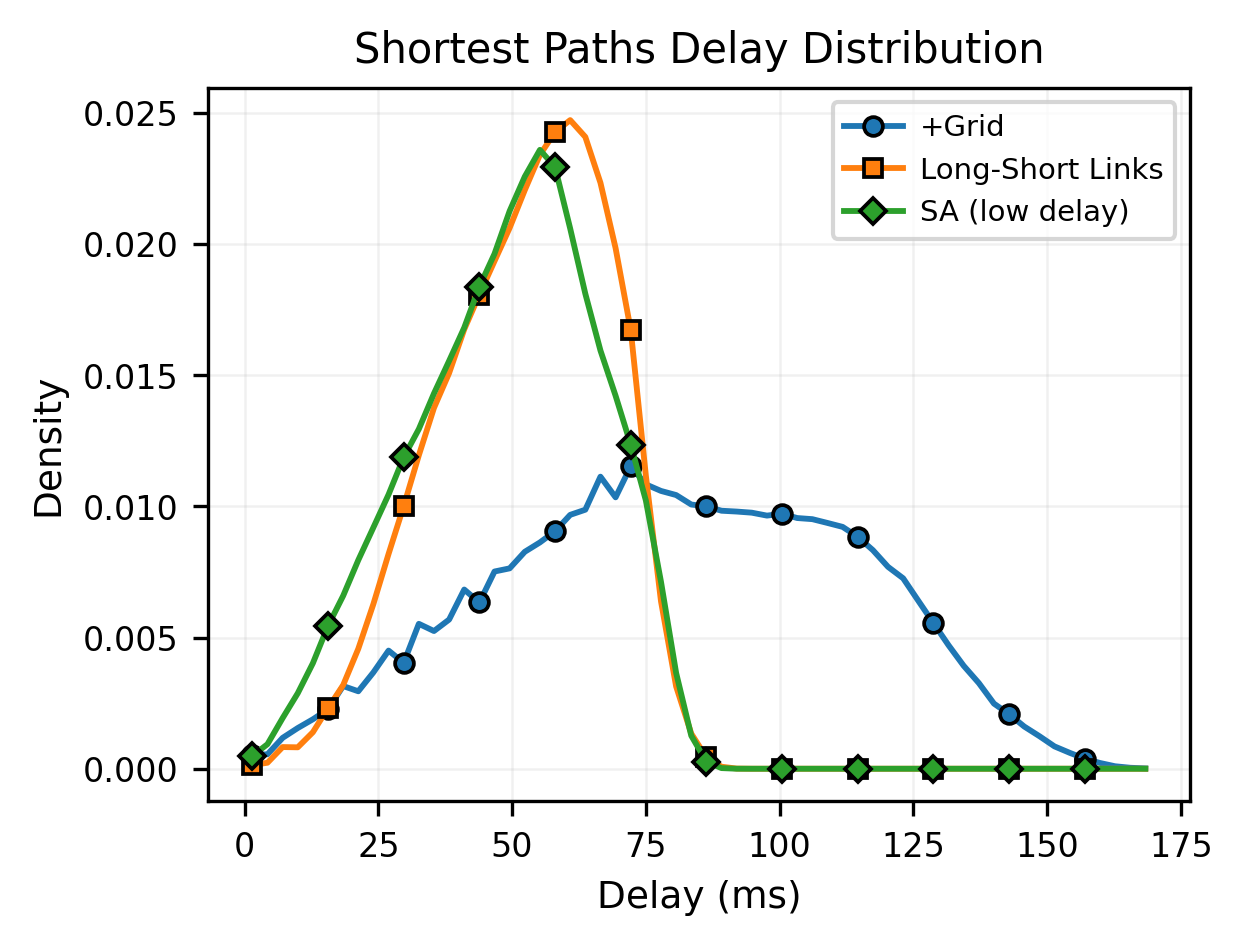}
    \caption{Delay distribution (ISL=4, 2024-10-02).}
    \label{fig:real_delay_dist_4}
  \end{subfigure}\hfill
  \begin{subfigure}{0.245\textwidth}
    \centering
    \includegraphics[width=\linewidth]{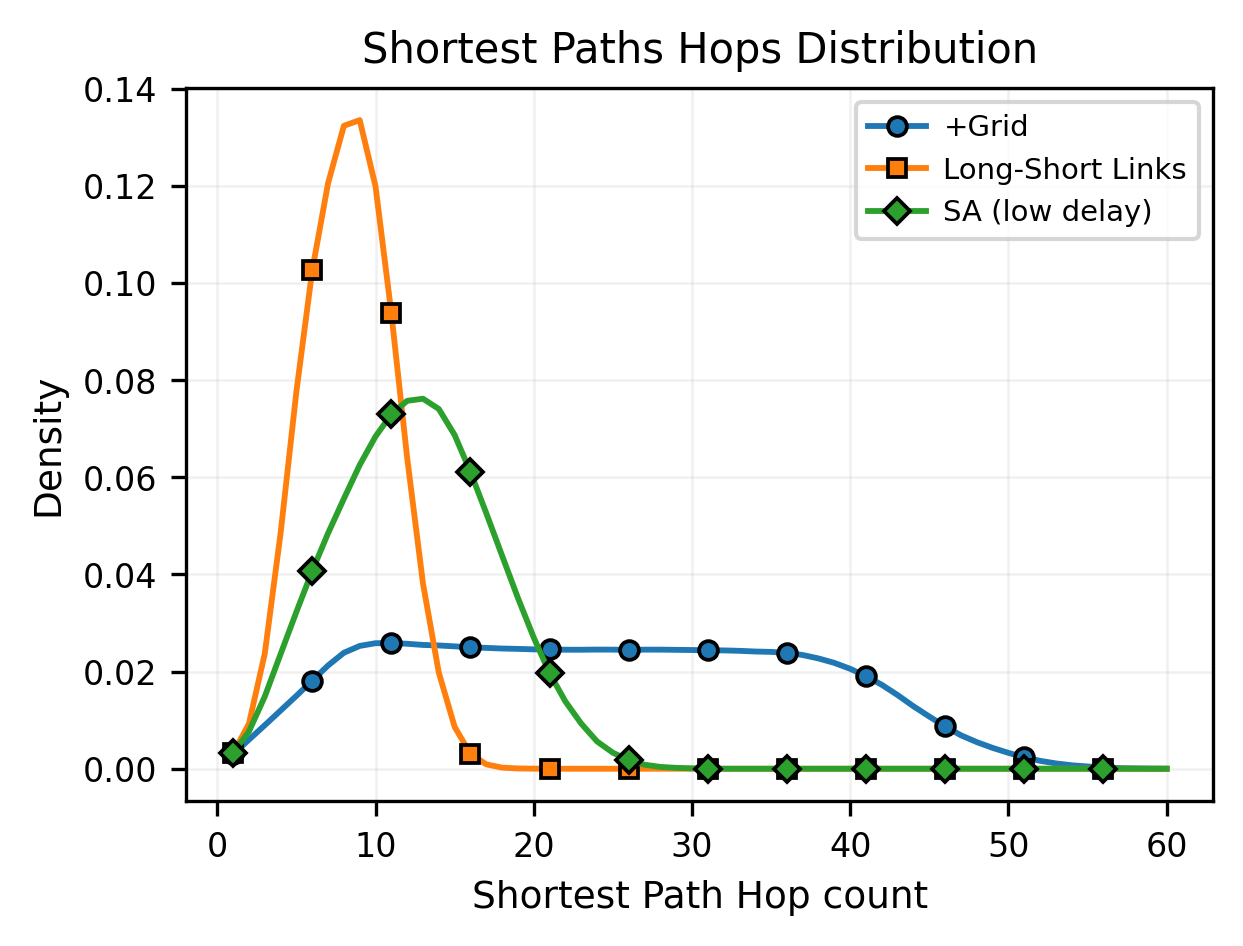}
    \caption{Hop Count distribution (ISL=4, 2024-10-02).}
    \label{fig:real_hop_dist_4}
  \end{subfigure}

  \vspace{0.35em}

  \begin{subfigure}{0.245\textwidth}
    \centering
    \includegraphics[width=\linewidth]{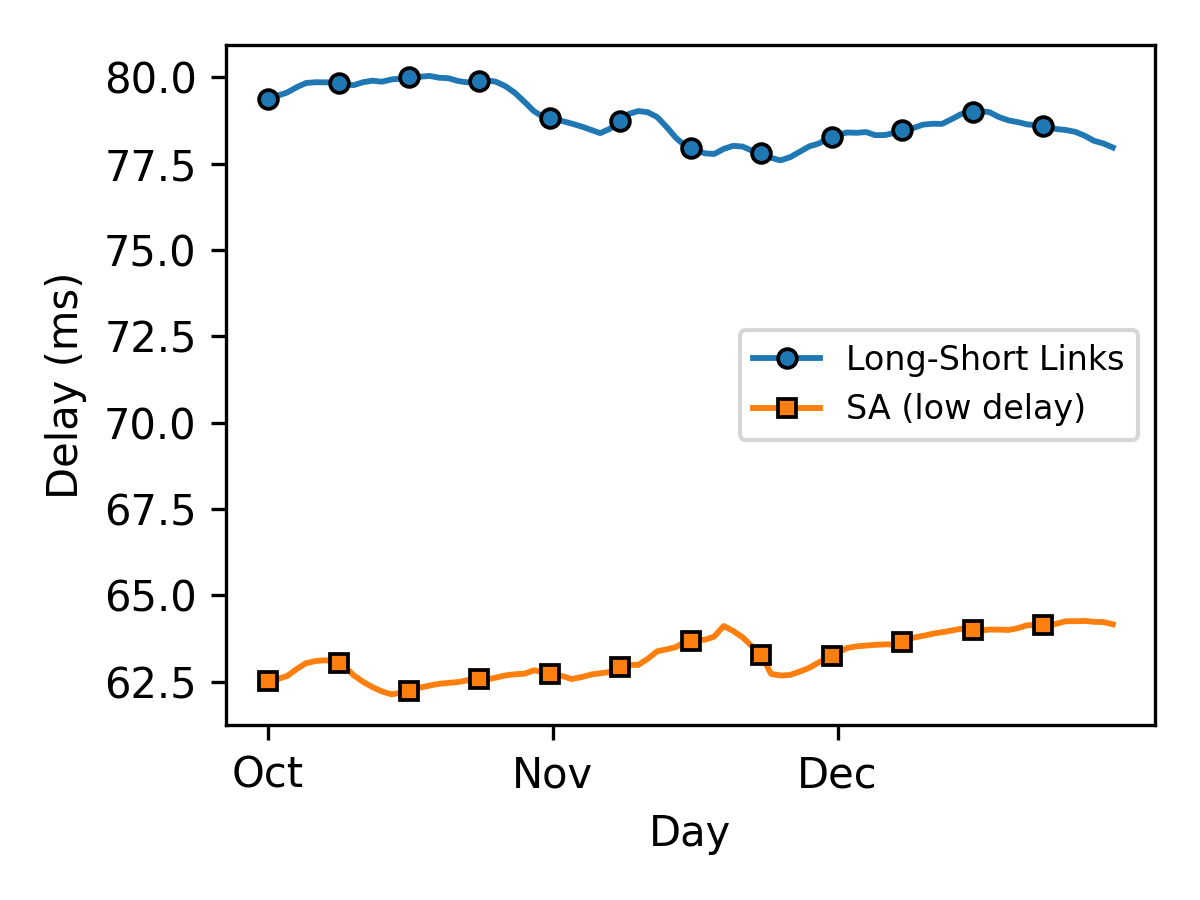}
    \caption{Delay over time (ISL=3).}
    \label{fig:real_delay_over_time_3}
  \end{subfigure}\hfill
  \begin{subfigure}{0.245\textwidth}
    \centering
    \includegraphics[width=\linewidth]{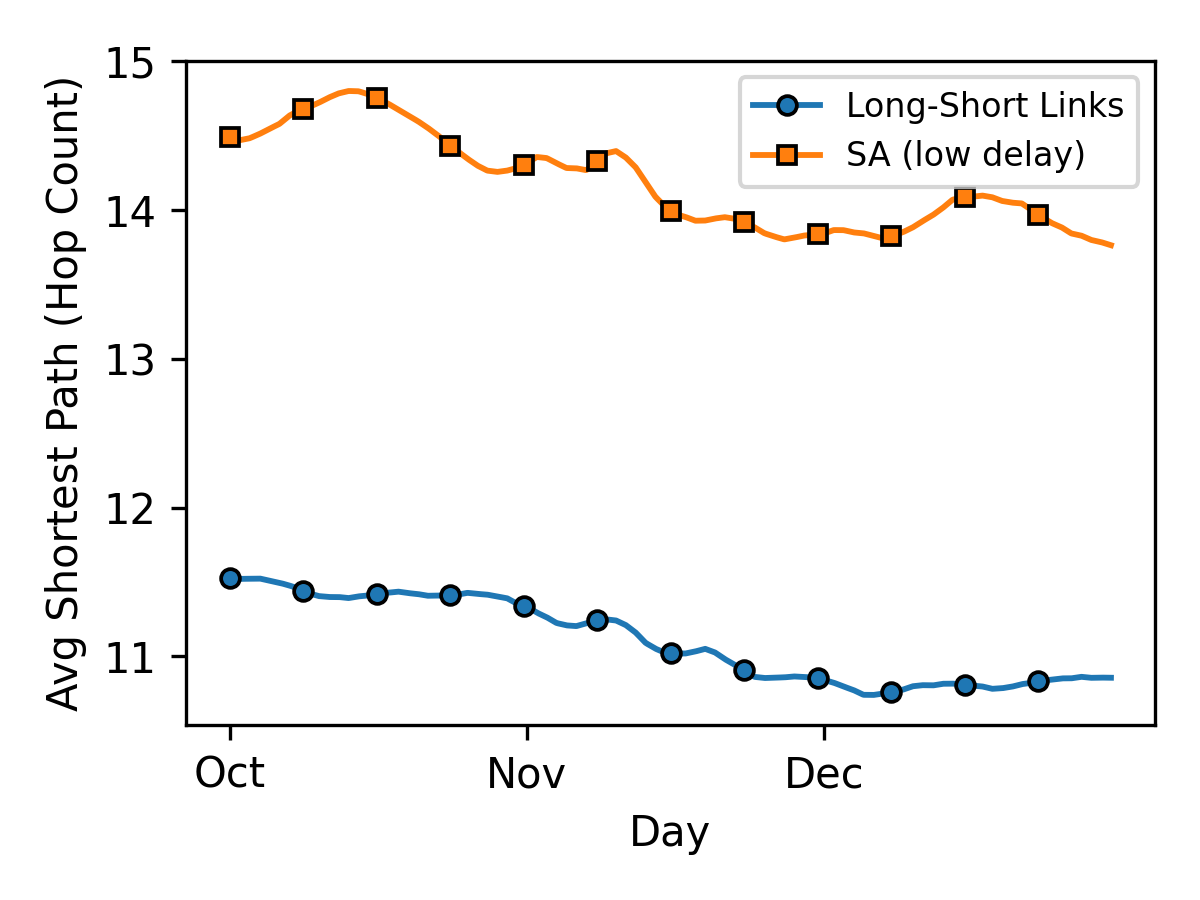}
    \caption{Hop count over time (ISL=3).}
    \label{fig:real_hop_over_time_3}
  \end{subfigure}\hfill
  \begin{subfigure}{0.245\textwidth}
    \centering
    \includegraphics[width=\linewidth]{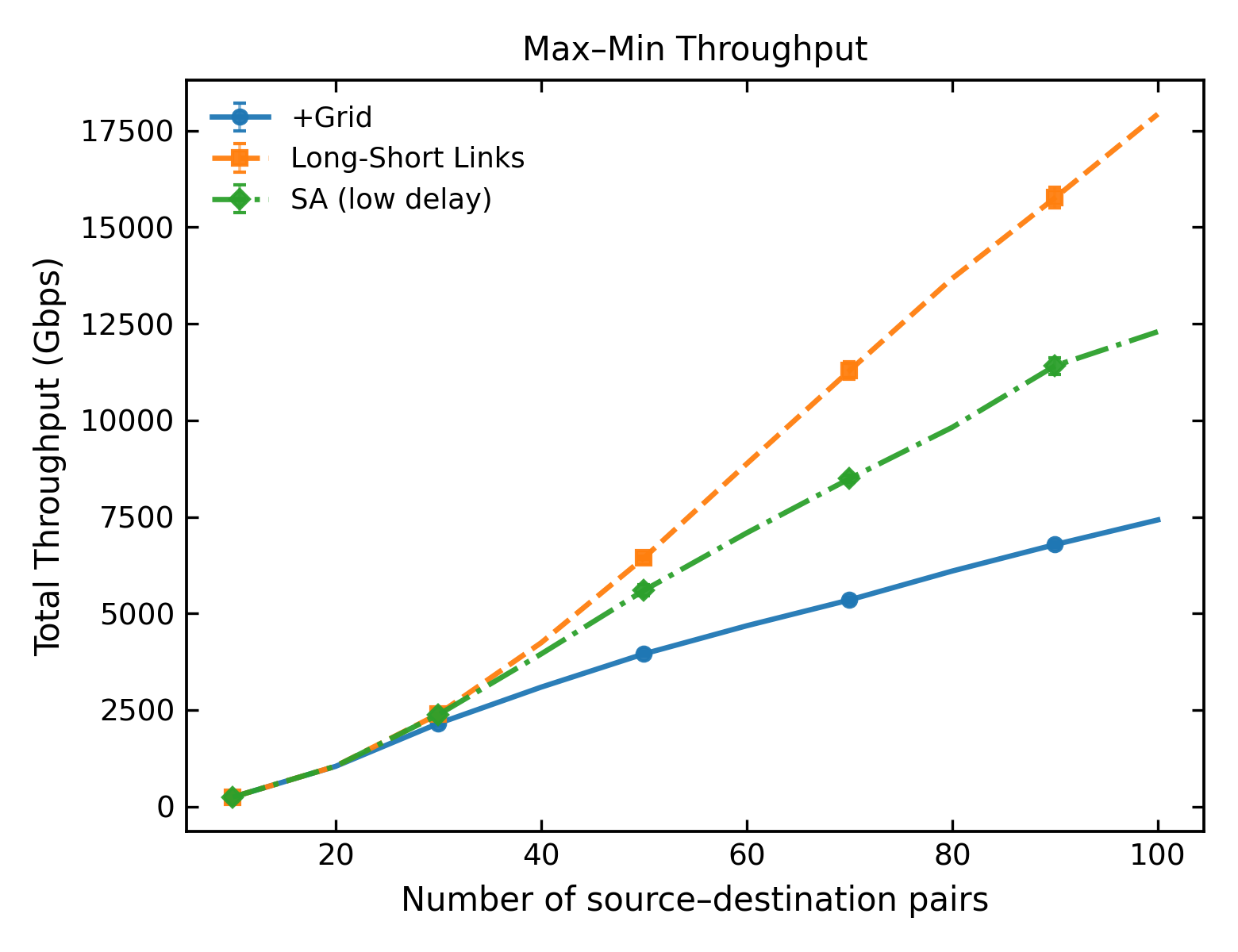}
    \caption{Throughput (ISL=4) (2024-10-02).}
    \label{fig:real_throughput_4}
  \end{subfigure}\hfill
  \begin{subfigure}{0.245\textwidth}
    \centering
    \includegraphics[width=\linewidth]{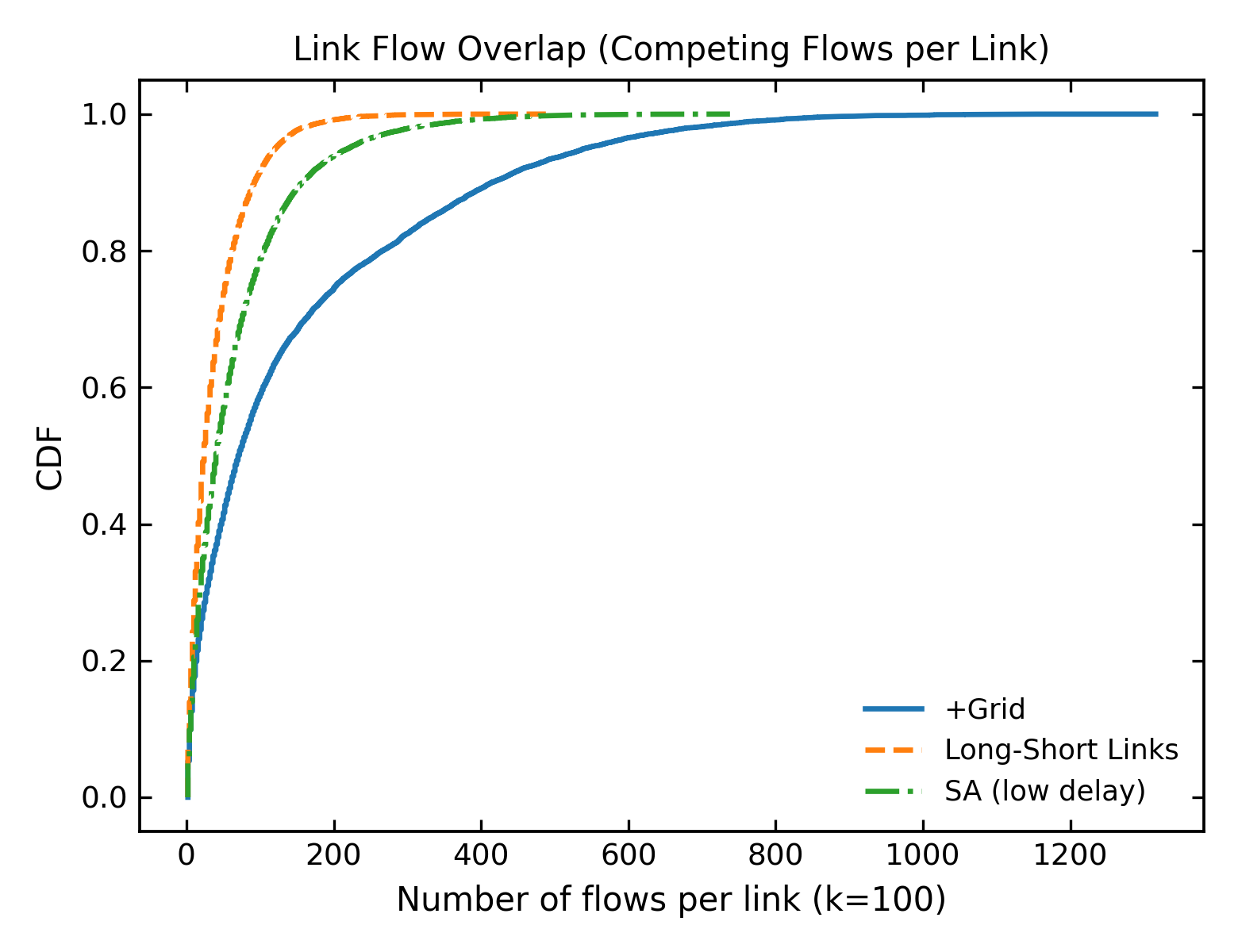}
    \caption{Flows per link for 100 src-dest pairs (ISL=4).}
    \label{fig:flows_per_link_real}
  \end{subfigure}

  \caption{Real Deployed Starlink Shell-1 (Oct--Dec 2024) evaluation.}
  \label{fig:real_compact_all}
\end{figure*}
{\textbf{SA Design Space and Surrogate objectives:}}
\label{subsubsec:pareto} To study the trade-off between average shortest-path delay and hop count, we sweep the surrogate objective weights and generate a family of SA designs over Starlink Shell-1. Specifically, we run SA over a grid $(\alpha_L,\alpha_U,\alpha_M) \in \{1,\dots,5\}^3$, yielding 125 feasible topologies. Figure~\ref{fig:pareto_all} visualizes this design space. We evaluate every topology by its average shortest-path delay and hop count, and then extract the Pareto front---designs for which no other topology achieves both lower delay and fewer hops. The Pareto front exhibits a smooth trade-off between delay and hop count: varying the surrogate weights moves the solution along this curve, allowing an operator to choose between lower delay and fewer hops. Using the same 125 designs, we examine how our surrogates correlate with end-to-end metrics. Figure~\ref{fig:len_vs_delay_corr_sub} shows that solutions with \emph{shorter mean link length} tend to achieve lower average shortest-path delay, supporting our delay surrogate. Figure~\ref{fig:long_vs_hop_corr_sub} shows that solutions with a \emph{higher fraction of long-range links} tend to achieve lower average shortest-path hop count, supporting our hop surrogate. We define a \emph{long-range} link as one whose endpoints are separated by more than three orbital planes.

\subsubsection{\textbf{Network capacity}}
\label{subsec:traffic_capacity}

A topology should not only yield short paths but also sustain high throughput under many concurrent flows. We evaluate throughput on \emph{Starlink Shell-1} using a max--min fair allocation model, which reveals bottlenecks under contention. To enable a meaningful comparison, we subject all topologies to identical traffic demands, routing, and link-capacity constraints, and estimate their achievable aggregate throughput via max--min fairness~\cite{max_min_tutorial}.

{\textbf{Max--min fairness:}}
Max--min fairness allocates bandwidth by maximizing the minimum rate across all flows. A flow’s rate can be increased only if doing so does not reduce the rate of another flow with an equal or smaller allocation. This criterion provides a principled and widely used basis for comparing how different topologies share limited link capacity under concurrent demand~\cite{data_networks}.

{\textbf{Methodology:}}
We evaluate network capacity on a synthetic Starlink Shell-1 configuration. We randomly select from 10 to 100 source--destination satellite pairs and assign each pair a traffic demand drawn from a Gaussian distribution in the range $[0.5, 5]$\,Gbps. Each inter-satellite link is modeled as full-duplex with a constant capacity of 100\,Gbps per direction, based on current hardware~\cite{tesat_scot80,starlink_report}. Traffic is routed using single shortest-path routing, and end-to-end throughput flow rates are computed using max--min fair bandwidth allocation. We then compute aggregate throughput across all flows. To reduce variance, we repeat each source--destination configuration 10 times and report the average aggregate throughput across all flows.

{\textbf{Findings:}} Figure~\ref{fig:throughput_shell1_syn} shows aggregate throughput on Synthetic Starlink Shell-1. We observe that designs that produce shorter hop-count paths achieve greater path diversity and higher throughput. \emph{LSL attains the highest throughput}, reaching up to 20~Tbps for 100 src-dest pairs, about $2\times$ of Motif and +Grid, which achieve around 10--11~Tbps. This gap arises from path concentration. Motif and +Grid route many source--destination pairs through a small set of shared links, leading to early congestion and reduced throughput. Figure~\ref{fig:flows_per_link_syn} illustrates this effect. Under Motif and +Grid, many flows traverse the same links, whereas LSL and SA-based designs distribute traffic more evenly across the network, reducing contention.

\emph{SA-based designs also perform strongly.} Among them, SA (low hop count) achieves the second-highest throughput, followed by SA (balanced) and SA (low delay), with all three outperforming Motif and +Grid. Lower-hop solutions naturally promote greater path diversity by reducing overlap among flows, which alleviates bottlenecks under max--min fairness.

Importantly, these gains are achieved \emph{without any explicit traffic engineering or multipath routing}. Even under single shortest-path routing, LSL and SA-based topologies induce substantially more diverse paths, which directly translates into higher throughput.


\subsection{Real Constellation Results}
We next evaluate on daily Starlink Shell-1 TLE snapshots from October--December 2024, capturing real deployment dynamics with time-varying satellite availability and link feasibility. We compute the topology from scratch on day~1 (2024-10-01). For each subsequent day, we update it using the incremental update algorithm. 

Due to partial deployment, structured baselines are generally not applicable to real Starlink Shell-1. 
Motif assumes a fully deployed, symmetric constellation and therefore does not apply under deployment asymmetry. 
+Grid also assumes symmetry. For comparison, we adapt it to partial deployments by connecting each satellite to its nearest available neighbors, allowing fewer than four ISLs when necessary.

For the 3-ISL setting, 3-ISL-Grid requires a strict alternating inter-plane pattern that cannot be realized with uneven plane populations. Instantiating it results in a disconnected graph, so we exclude it from evaluation.

\subsubsection{\textbf{Path Quality}}
\label{subsec:real_shell1}

For each day of 3-month period, we  measure average shortest-path delay and hop count. The results are shown in Figure~\ref{fig:real_compact_all}.

{\textbf{Findings:}} Across daily Starlink Shell-1 snapshots (Oct--Dec 2024), our methods remain consistently strong despite deployment dynamics.

Under the 4-ISL setting, both of our methods substantially outperform the +Grid baseline. +Grid yields an average delay of 89.0\,ms with 23.4 hops. LSL reduces delay to 52.9\,ms ($40\%$ reduction) and hop count to 8.2 ($65.0\%$ reduction). SA (low-delay) achieves the lowest delay at 49.3\,ms ($45\%$ reduction) while keeping hop count moderate at 12.0 ($49\%$ reduction).

For 3-ISL configurations, SA adapts well to the tighter degree budget and continues to achieve favorable delay--hop trade-offs, with roughly 60\,ms average delay and about 10--11 hops. LSL maintains relatively low hop counts, but incurs higher delay (77--80 ms) under the stricter 3-ISL limit, consistent with the reduced availability of short links. Even under these constraints, both approaches deliver substantially lower hop counts than +Grid, which remains in the 20s even when allowed a higher limit of 4 ISLs.

\subsubsection{\textbf{Network Capacity}} 
\label{sec:traf_cap_real}
Using the same methodology as in Section~\ref{subsec:traffic_capacity} for the synthetic shell, we compute throughput on the first day of the three-month snapshot period.

{\textbf{Findings:}}
Figure~\ref{fig:real_throughput_4} shows aggregate throughput on real Starlink deployment snapshots under an ISL limit of 4.

\emph{LSL and SA-based designs consistently achieve high throughput} and closely mirror the trends observed on synthetic shells. In particular, LSL sustains substantially highest aggregate throughput, reaching up to 17.5~Tbps with 100 source--destination pairs, compared to 7.5~Tbps for the +Grid baseline, representing nearly a $2.3\times$ improvement. SA achieves about 12.5~Tbps throughput, nearly a $1.7\times$ improvement over +Grid.

As in the synthetic setting, the key factor is path diversity. Figure~\ref{fig:flows_per_link_real} shows that under +Grid, many flows are concentrated on a small number of links, which leads to early congestion due to limited alternative paths. In contrast, LSL- and SA-based topologies spread traffic more evenly across the network, with far fewer flows sharing the same links.

As discussed before, these throughput gains arise \emph{without any explicit traffic engineering}. Even under single shortest-path routing, LSL and SA induce more diverse end-to-end paths through their topology alone, which directly translates into higher throughput and improved congestion resilience in real Starlink deployments.

\subsection{Robustness}
\label{subsec:robustness}

Real constellations evolve over time due to launches, de-orbiting, and temporary satellite unavailability. Recomputing a topology from scratch each day can induce substantial link churn. To mitigate this, we use an \emph{incremental update} procedure for both LSL and SA that preserves as much of the previous day's topology as possible.


To evaluate the effectiveness of our incremental update procedure, we conduct two experiments: an analysis of day-to-day link breakages, and an ablation study of robustness under shell growth and shrinkage.

\subsubsection{\textbf{Link breakages}}
\label{subsec:link_breakages}

\begin{figure}[h]
  \centering
  \captionsetup[subfigure]{font=scriptsize}

  \begin{subfigure}{0.47\linewidth}
    \centering
    \includegraphics[width=\linewidth]{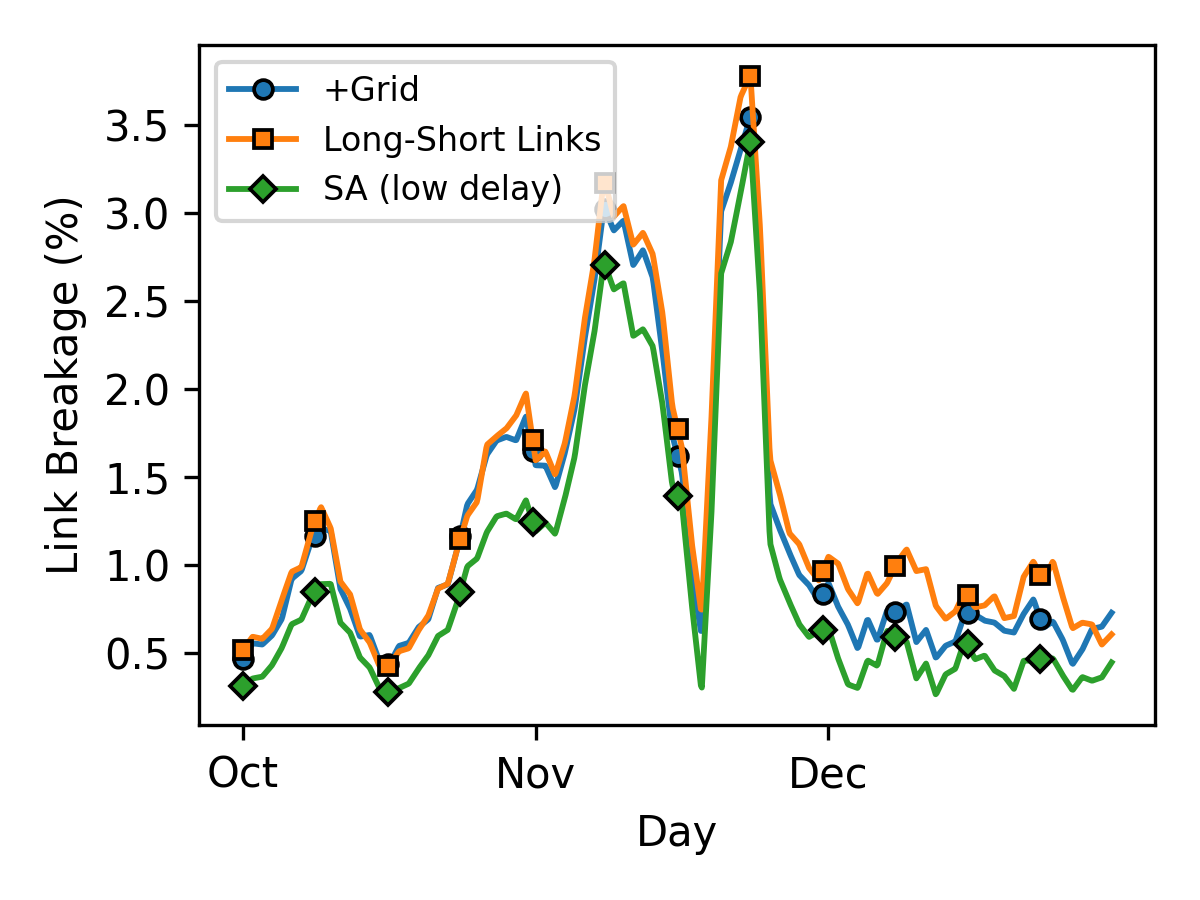}
    \caption{Link breakages with 4 ISLs per satellite.}
    \label{fig:linkfail_4isl}
  \end{subfigure}
  \hfill
  \begin{subfigure}{0.47\linewidth}
    \centering
    \includegraphics[width=\linewidth]{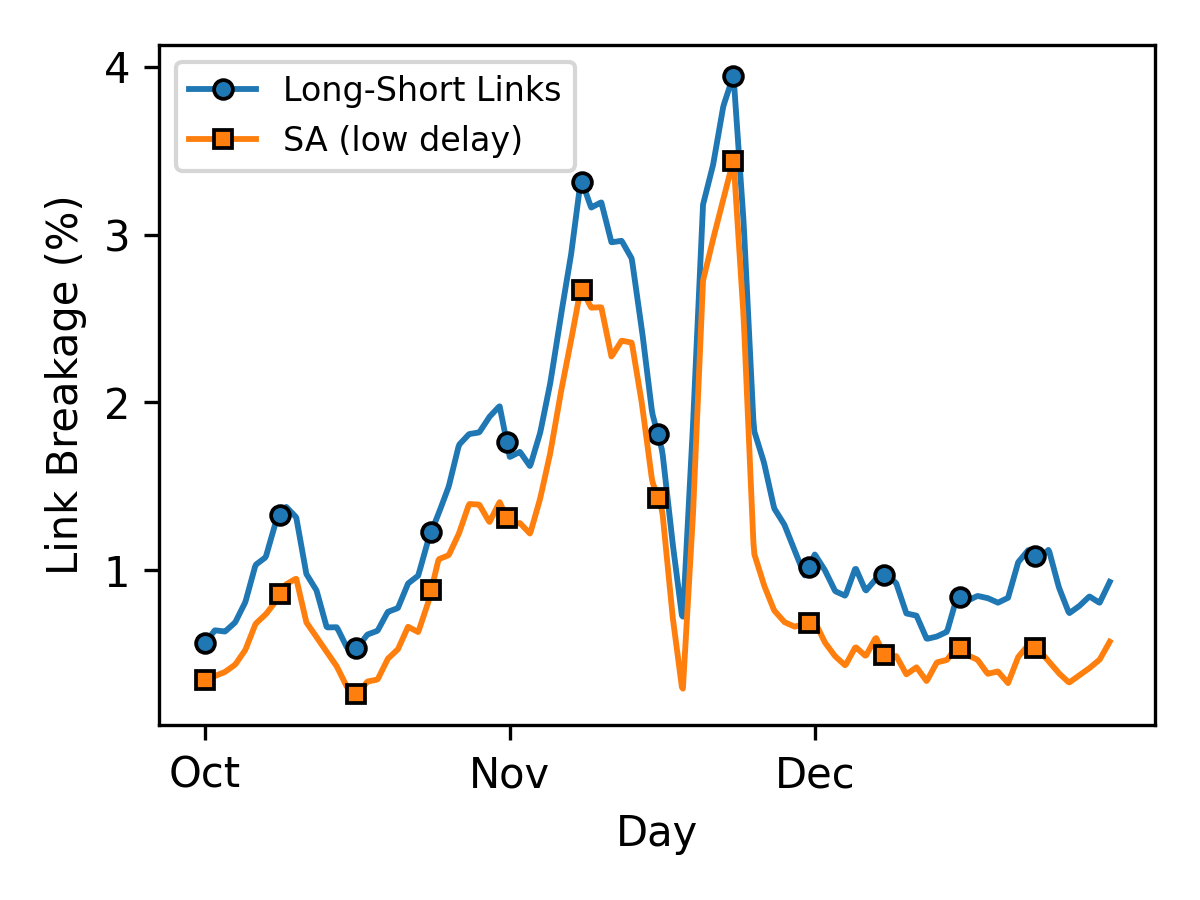}
    \caption{Link breakages with 3 ISLs per satellite.}
    \label{fig:linkfail_3isl}
  \end{subfigure}

  \caption{Link Breakages in Real Starlink Shell-1 (Oct--Dec 2024).}
  \label{fig:real_link_breakages}
\end{figure}

Figure~\ref{fig:real_link_breakages} reports day-to-day ISL breakages on real Starlink Shell-1 snapshots. Breakages arise from satellite repositioning (which changes link feasibility) and node turnover (launches and de-orbits). Our designs have breakage rates comparable to the +Grid baseline: SA averages \(1.0\%\), LSL averages \(1.3\%\), and +Grid averages \(1.2\%\). This indicates that our performance gains do not come at the cost of increased churn. This low breakage rate is attributed to the usage of stable links.

\subsubsection{\textbf{Robustness Under Shell Growth and Shrinkage}}
\label{subsec:robustness_incremental}

\begin{figure}[h]
  \centering
  \captionsetup{font=footnotesize}
  \captionsetup[subfigure]{font=scriptsize}

  \begin{subfigure}{0.7\linewidth}
    \centering
    \includegraphics[width=\linewidth]{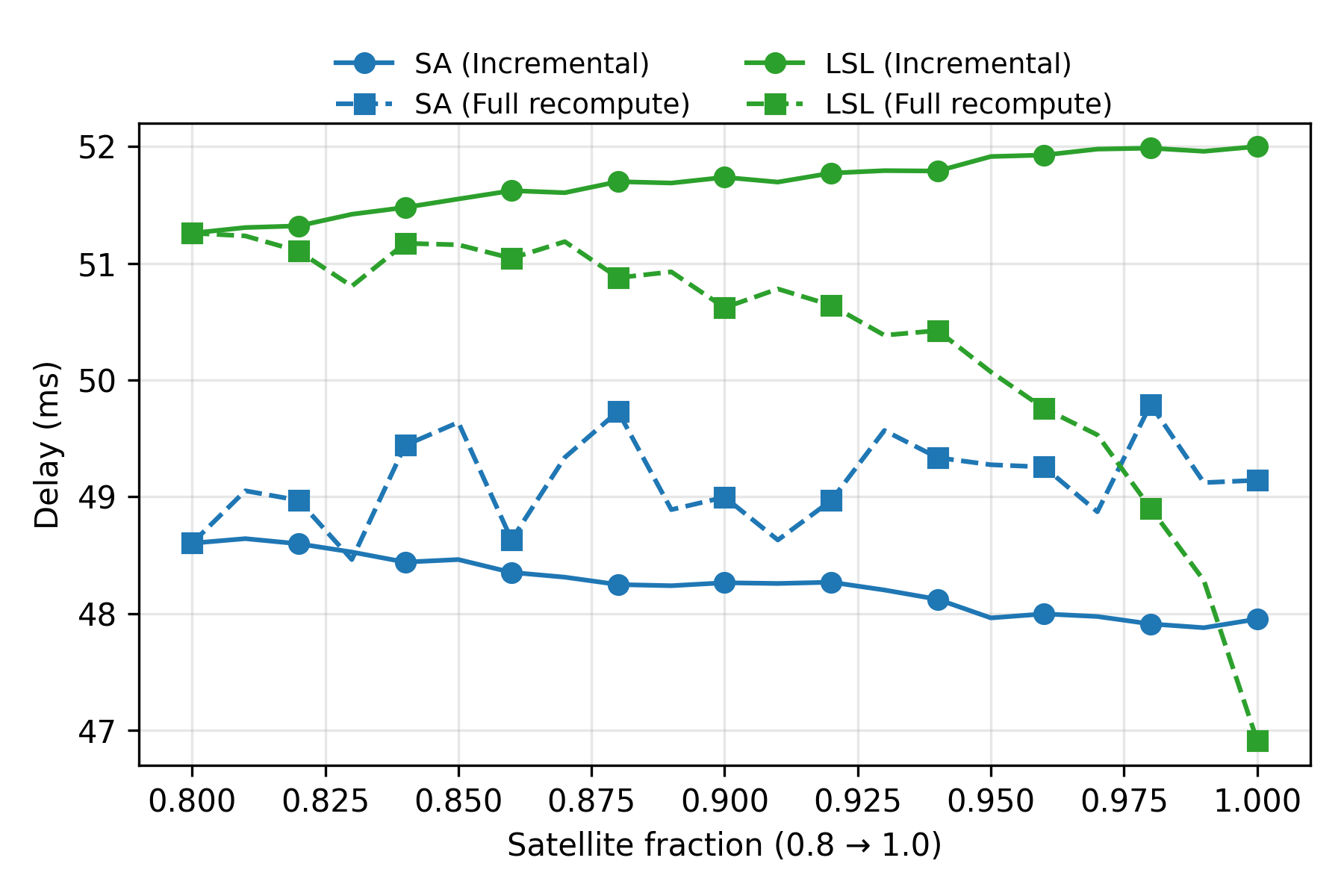}
    \caption{Shell growth: 80\% $\rightarrow$ 100\%.}
    \label{fig:robustness_growth}
  \end{subfigure}

  \vspace{0.6em}

  \begin{subfigure}{0.7\linewidth}
    \centering
    \includegraphics[width=\linewidth]{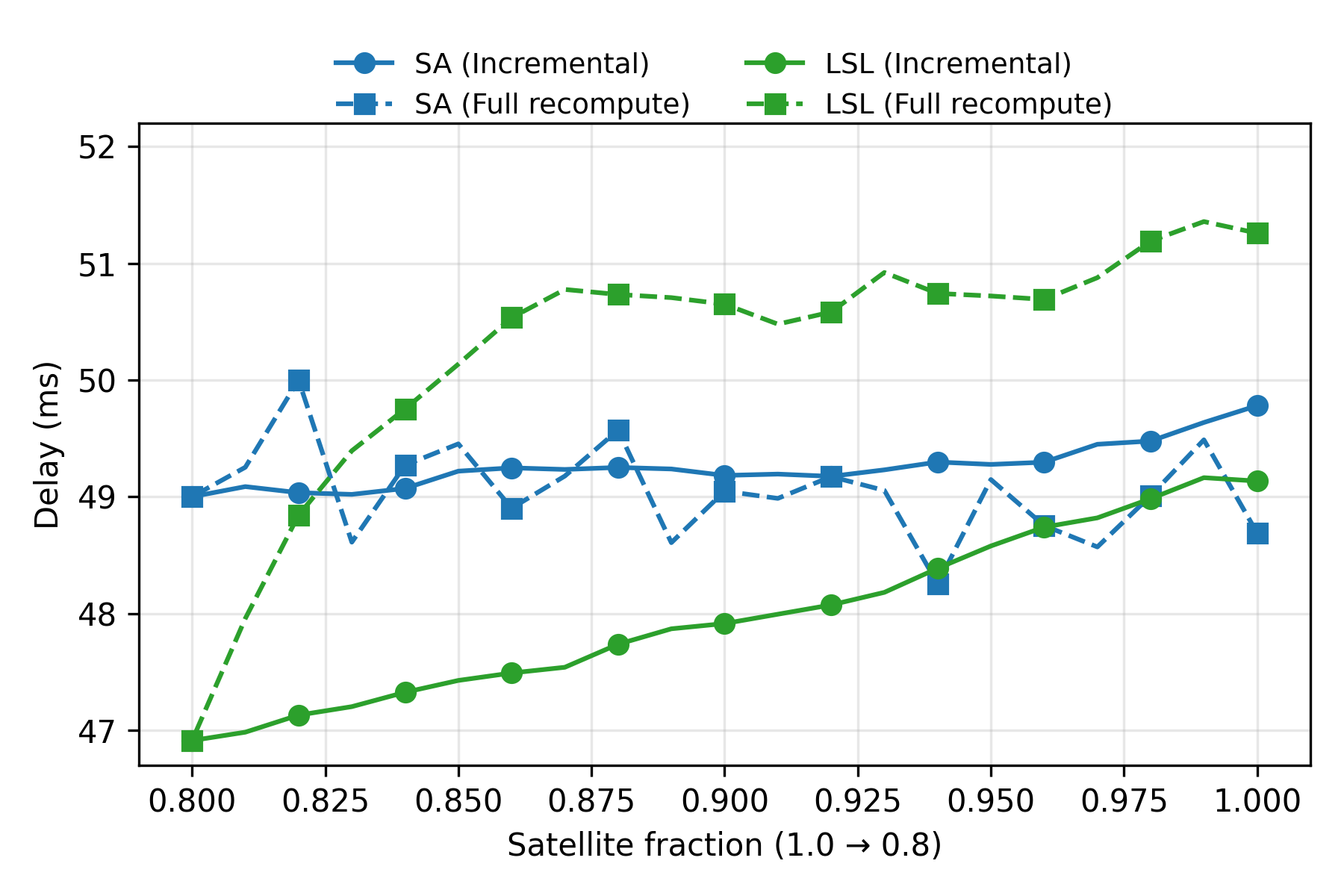}
    \caption{Shell shrinkage: 100\% $\rightarrow$ 80\%.}
    \label{fig:robustness_shrink}
  \end{subfigure}

  \caption{Delay Comparison: Incremental updates vs.\ full recomputation under shell growth and shrinkage scenarios.}
  \label{fig:robustness_incremental}
\end{figure}

 Real constellations change over time: new satellites are launched and integrated, while others are de-orbited or temporarily unavailable. To isolate the effect of such node turnover on our topology design methods, we run controlled \emph{growth} and \emph{shrinkage} experiments on \emph{synthetic Starlink Shell-1}. Starting from a partially populated shell, we vary the active satellite set by 1\% per day under two scenarios:
\begin{itemize}[noitemsep,topsep=0pt]
  \item \textbf{Growth:} 80\% $\rightarrow$ 100\% of full size (add 1\% satellites/day).
  \item \textbf{Shrinkage:} 100\% $\rightarrow$ 80\% of full size (remove 1\% satellites/day).
\end{itemize}
On each day, for both LSL and SA, we compare an \emph{incrementally updated} topology (preserving prior links when feasible) against a topology \emph{recomputed from scratch} on that day's active set. We then measure the average shortest-path delay and hop count for both methods (Figures~\ref{fig:robustness_incremental} and \ref{fig:robustness_incremental_hop}).

{\textbf{Findings:}} Figure~\ref{fig:robustness_incremental} compares shortest-path delay under incremental updates against full recomputation for both shell growth and shrinkage. 
\emph{SA is robust to node turnover:} in both scenarios, incremental delay closely tracks full recomputation with only minor deviations, and the curves remain smooth, indicating that the update procedure in Section~\S\ref{subsec:incremental_updates} preserves path quality without inducing unnecessary churn.

\emph{LSL remains stable, with small systematic offsets:} incremental delay stays within \(\approx 1\)–\(5\)~ms of recomputation. Under growth it is slightly higher, while under shrinkage it is slightly lower. This asymmetry is driven by initialization: in shrinkage, LSL starts from a fully deployed topology and retains efficient persistent links, whereas in growth it starts from a smaller shell and early suboptimal choices can carry forward as the network expands.

Hop count is robust: Appendix Figure~\ref{fig:robustness_incremental_hop} shows that, for both LSL and SA, the hop count under incremental updates closely mirrors the hop count under full recomputation.

Moreover, if the incrementally updated topology and recomputed one ever diverges significantly over long horizons, an operator can periodically realign by switching to a fully recomputed topology.


\section{Discussion}
\label{sec:discussion}

Our designs use \emph{stable links}, defined as inter-satellite links that remain feasible over a full orbital cycle. Using transient links, such as those between oppositely moving satellites, can potentially give better performance by creating shorter paths and reducing end-to-end delay. However, using such links introduces high churn, requires appropriate hardware, and complicates incremental topology maintenance. This trade-off motivates our focus on stable links.

We further validate this choice experimentally using a synthetic Starlink Shell-1 configuration. We compare two idealized cases with no ISL degree constraints: (i) topologies using only stable links, and (ii) topologies using all feasible links, including transient ones. The average shortest-path delay is approximately 38\,ms with only stable links, compared to 36\,ms when all links are allowed. This small gap shows that stable links capture most of the path-quality benefits of the full link set, while offering far greater robustness and predictability.

In future work, we propose jointly considering topology design, routing, and traffic engineering, which are tightly coupled in LEO networks. While co-optimizing them could yield the best performance, deployment constraints (limited degree, time-varying feasibility, evolving constellations) make this difficult. 




\section{Related Works}
\label{sec:related_works}

\paragraph{Works in LEO topology design.}
Prior work has proposed several approaches for designing ISL topologies in LEO. These works typically assume a largely static and symmetric constellation, overlooking important real-world constraints such as time-varying link and node availability.


\textit{Motif}~\cite{motif}.
Motif selects two-link motifs from a repeating pattern space defined over an ideal, symmetric constellation. So in practice, it does not apply to operational LEO constellations because of asymmetry. It also assumes a structure of 4 ISLs per satellite.
whereas our designs apply to both 3-ISL and 4-ISL settings.

\textit{DoTD}~\cite{ron2025timedependent}.
DoTD uses dynamic programming to design topologies over a time-dependent edge graph, jointly minimizing latency, network capacity, and link churn. However, it does not account for changing node availability, which, as we show, is a persistent factor in real deployments.
Moreover, it treats all feasible links uniformly, without distinguishing highly stable links from marginal ones, and can therefore incur significant churn.

\textit{AlphaSat}~\cite{mcts}.
AlphaSat uses Monte Carlo Tree Search (MCTS) combined with a neural network to design LEO topologies. It only optimizes over a single static snapshot, without accounting for link stability or modeling incremental updates under evolving node availability. Its runtime is also prohibitive for mega-constellation and long-horizon evaluations, scaling as \(O(N^{2} N_{\max} M^{3})\), where \(M\) is the number of satellites, \(N\) is the number of cities, and \(N_{\max}\) denotes the per-satellite ISL limit.

In addition to the above works, prior work has studied the minimum achievable hop count in idealized satellite constellations~\cite{minhop_count} and examined how orbital and constellation parameters influence network performance in fully deployed models~\cite{TopologyDesignParameters}.

In contrast to these approaches, we focus on topology design for real deployments, where node availability, link availability, and link length vary over time. To address these challenges, we propose two approaches that operate directly on evolving constellations. Our designs are efficient at constellation scale and can be recomputed over long horizons, as demonstrated in our evaluation Section~\S\ref{sec:evaluation}.


\paragraph{LEO Measurement Works}
Recent work has measured and characterized real LEO networks using both dedicated terminals and Internet-exposed vantage points, providing empirical views of Starlink latency and routing dynamics~\cite{lens,democratizeLeo}. Other studies examine resilience and performance under disruptions and operational dynamics, including failures, reentries, and orbit-maintenance maneuvers~\cite{resillienceLEO,Oliveira_2025,CollisionAvoidanceStarlink}.

\section{Ethics Statement}
\label{sec:ethics}

This work does not involve human subjects, personal data, or interactions with real users. All experiments are conducted using simulations and publicly available datasets. We do not perform any measurements that interfere with operational networks or deployed systems. As a result, this study does not raise ethical concerns related to privacy, consent, or harm.

\section{Conclusion}
\label{sec:conclusion}

Most prior work on LEO topology design assumes fully deployed, static constellations with persistent links. These assumptions do not hold in real deployments. We address this gap by proposing two topology design methods that explicitly operate under deployment dynamics. Our evaluation on realistic constellations shows that these approaches consistently outperform existing baselines, achieving substantially lower hop count and delay, and higher throughput.

\bibliographystyle{ACM-Reference-Format}
\bibliography{citations}

\clearpage

\appendix

\appendix
\section{Additional Figures}

\begin{figure}[H]
  \centering
  \captionsetup{font=footnotesize}
  \captionsetup[subfigure]{font=scriptsize}

  \begin{subfigure}{0.7\linewidth}
    \centering
    \includegraphics[width=\linewidth]{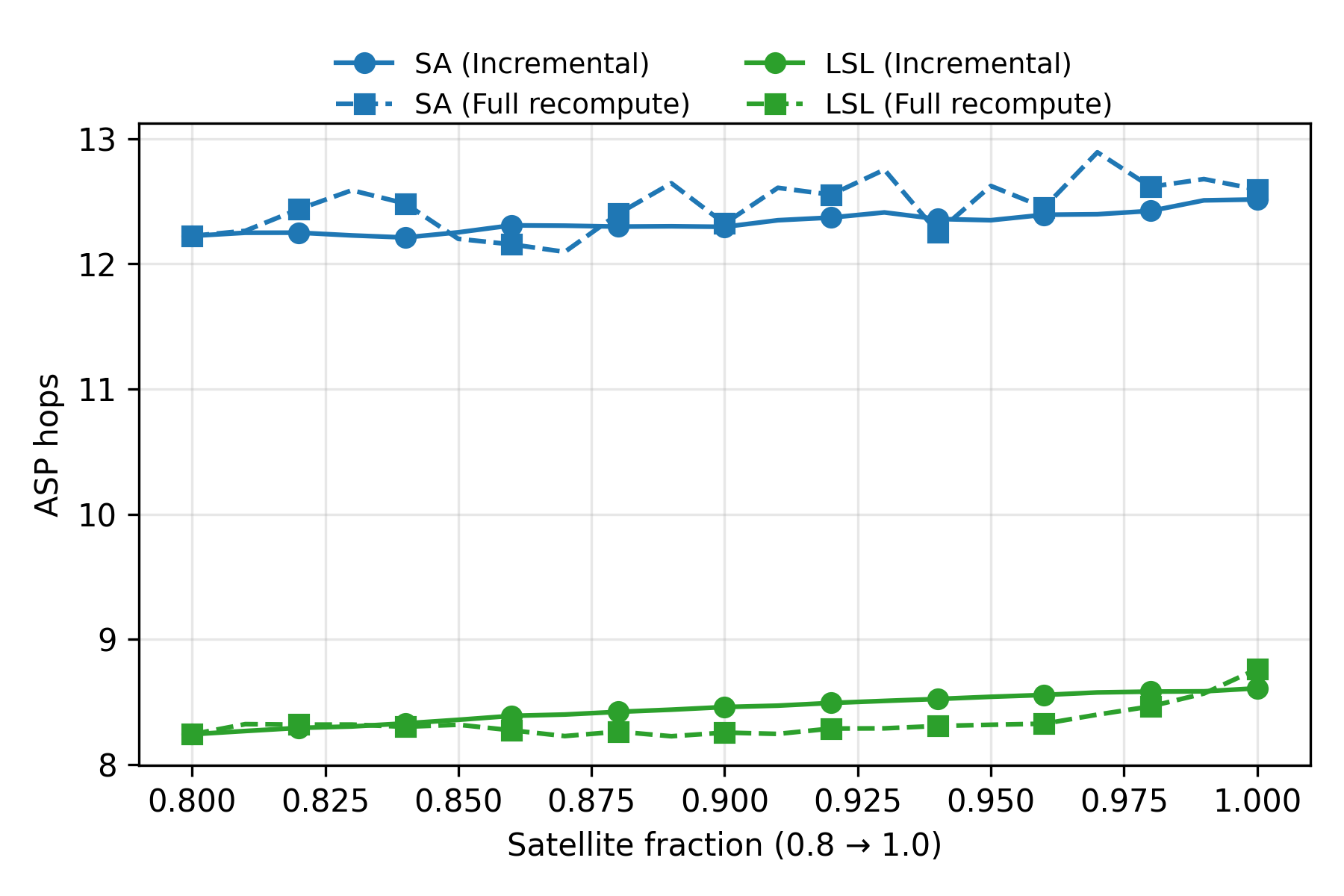}
    \caption{Shell growth: 80\% $\rightarrow$ 100\%.}
    \label{fig:robustness_growth_hops}
  \end{subfigure}

  \vspace{0.6em}

  \begin{subfigure}{0.7\linewidth}
    \centering
    \includegraphics[width=\linewidth]{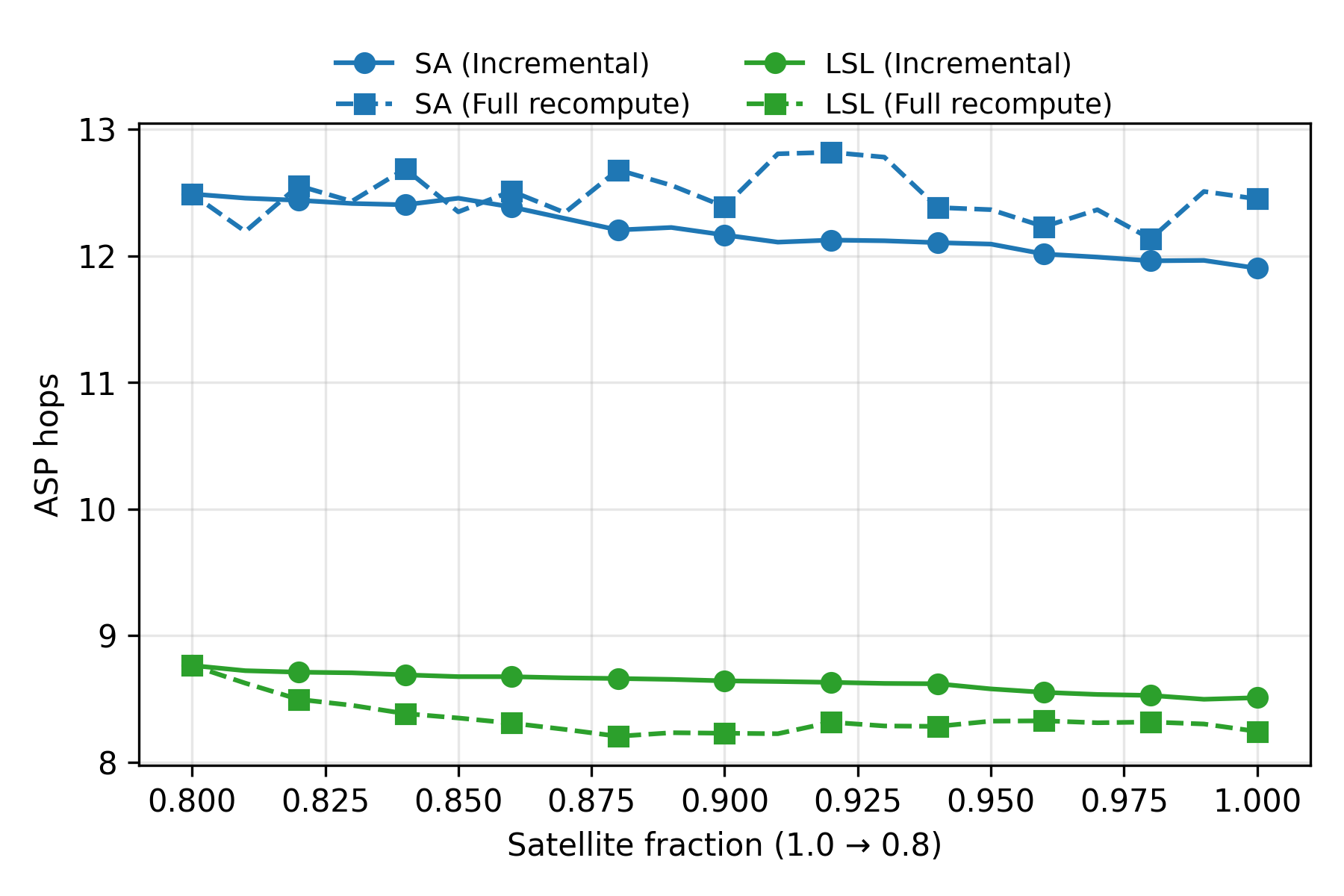}
    \caption{Shell shrinkage: 100\% $\rightarrow$ 80\%.}
    \label{fig:robustness_shrink_hops}
  \end{subfigure}

  \caption{Hop Count Comparison: Incremental updates vs.\ full recomputation under shell growth and shrinkage scenarios.}
  \label{fig:robustness_incremental_hop}
\end{figure}

\section{Parameters}
\begin{table}[t]
\centering
\small
\begin{adjustbox}{max width=\linewidth}
\begin{tabular}{llcccccc}
\hline
Constellation 
& Method 
& $D$ 
& Topology Type 
& $(\alpha_L,\alpha_U,\alpha_M)$ 
& Init. Iters 
& Incr. Iters \\
\hline
Starlink Shell~1 (real + synthetic) 
& LSL 
& 9 
& -- 
& -- 
& -- 
& -- \\

Starlink Shell~1 (real + synthetic) 
& SA  
& -- 
& Low delay 
& $(4,1,1)$ 
& 200k 
& 100k \\

Starlink Shell~1 (real + synthetic) 
& SA  
& -- 
& Low hop   
& $(1,2,5)$ 
& 200k 
& 100k \\

Starlink Shell~1 (real + synthetic) 
& SA  
& -- 
& Balanced  
& $(5,3,2)$ 
& 200k 
& 100k \\
\hline
Amazon Kuiper Shell (synthetic) 
& LSL 
& 4 
& -- 
& -- 
& -- 
& -- \\

Amazon Kuiper Shell (synthetic) 
& SA  
& -- 
& Low delay 
& $(4,2,2)$ 
& 200k 
& 100k \\

Amazon Kuiper Shell (synthetic) 
& SA  
& -- 
& Low hop   
& $(2,5,3)$ 
& 200k 
& 100k \\

Amazon Kuiper Shell (synthetic) 
& SA  
& -- 
& Balanced  
& $(3,2,2)$ 
& 200k 
& 100k \\
\hline
\end{tabular}
\end{adjustbox}
\caption{Parameter settings for LSL and SA across constellations. 
LSL uses maximum orbital separation parameter $D$, while SA uses surrogate objective weights 
$(\alpha_L,\alpha_U,\alpha_M)$ and fixed iteration budgets 
(200k initial computation, 100k incremental update).}
\label{tab:param_settings}
\end{table}

\end{document}